\documentclass[showpacs,preprintnumbers,amsmath,amssymb,superscriptaddress]{revtex4-1}
\usepackage{ulem}
\usepackage[dvips]{graphicx}
\usepackage{color}
\usepackage{bm}
\numberwithin{equation}{section}
\begin{document}
\title{Calculable Microscopic Theory for $^{12}$C($\alpha$, $\gamma$)$^{16}$O Cross Section near Gamow Window II }
\author{Y. Suzuki}
\affiliation{Department of Physics, Niigata University, Niigata 950-2181, Japan}
\affiliation{RIKEN Nishina Center, Wako 351-0198, Japan}
\email{y.suzuki@emeritus.niigata-u.ac.jp}
\date{\today}

\begin{abstract}
\noindent
{\bf Abstract} A microscopic approach to the 
$^{12}$C$(\alpha, \gamma)^{16}$O radiative-capture reaction near the Gamow window has been proposed by Y. Suzuki, Few-Body Syst. {\bf 62}, 2 (2021). The important ingredients of the approach include the following: (1) The states of 
$^{12}$C and $^{16}$O relevant to the reaction are described by fully microscopic 3\,$\alpha$-particle and 4\,$\alpha$-particle configurations. (2) The isovector electric dipole transition is accounted for through the isospin impurity of the 
constituent $\alpha$-particles. (3) The relative motion among the $\alpha$-particles is expanded in terms of correlated-Gaussian basis functions. 
A calculation of the radiative-capture cross section demands double angular-momentum projections, that is, the angular momentum of  $^{12}$C consisting of 3 $\alpha$-particles and the orbital angular momentum for $^{12}$C$-\alpha$ relative motion. Advancing the previous formulation based on the single angular-momentum projection, I carry out the double projection and present all the formulas needed for the cross section calculation. 
\end{abstract}
\maketitle

\section{Introduction}

The radiative-capture reaction $^{12}$C($\alpha,\gamma$)$^{16}$O  is one
of the most important reactions in astrophysics. The 
helium-burning phase is essentially governed by the triple-$\alpha$ reaction, 
$\alpha \alpha(\alpha,\gamma)^{12}$C, and the $^{12}$C($\alpha,\gamma$)$^{16}$O 
reaction. Their reaction rates 
determine the ratio of $^{12}$C and $^{16}$O at the end of the helium-burning 
phase. Consequently, the $^{12}$C($\alpha,\gamma$)$^{16}$O reaction influences
strongly the subsequent nucleosynthesis processes for massive stars. The measurement of  the $^{12}$C($\alpha,\gamma$)$^{16}$O cross section at low energies is blocked by 
the Coulomb barrier. It has reached so far up to $E_{\rm c.m.}\approx 1$ MeV~\cite{deboer17}, which is too high compared to the Gamow window,  $E_{\rm c.m.}\approx 0.3$ MeV. Predicting the cross section theoretically has been desired.  

In a previous paper~\cite{suzuki21}, I presented a microscopic theory to study the $^{12}$C($\alpha,\gamma$)$^{16}$O radiative-capture reaction.  A basic assumption was that the relevant states of both $^{12}$C and $^{16}$O are all respectively described by $3\,\alpha$ and $4\,\alpha$ microscopic cluster models where the wave function of $\alpha$-particle is assumed to be  $(0s)^4$ configuration. The radiative-capture reaction at $E_{\rm c.m.}\approx 0.3$ MeV  takes place through electric dipole ($E1$) and electric quadrupole ($E2$) transitions. Because 
both of $^{12}$C and $\alpha$-particle are isospin zero nuclei, the $E1$ transition is forbidden in the leading-order of the long wavelength approximation.  The most crucial in the $^{12}$C+$\alpha$ radiative-capture calculation is therefore how to describe the $E1$ process, that is, how to take into account isospin-impurity components in the relevant nuclei. 
So far very few studies have confronted this problem directly. An exception is 
the microscopic calculation by $^{12}$C$+\alpha$ 
two-cluster model~\cite{pdesc87,pdesc87b}, where the isospin mixing in $^{16}$O was 
taken into account by the coupling to the $^{15}$N$+p$ and $^{15}$O$+n$ 
channels. 

I proposed in Ref.~\cite{suzuki21} an alternative way of introducing the isospin impurity, the distortion of $\alpha$-particle from the pure $T=0$ state. Its isospin-impurity  configuration has $L^{\pi}=0^+, S=0, T=1$ and  2$\hbar \omega$ excitation from the $(0s)^4$ configuration. The distortion of 
$\alpha$-particle of this type was the key to account for the $\beta$-decay of the ground state of $^9$Li to $^9$Be~\cite{arai96}.  
The mixing amplitude of the isospin-impurity component is estimated using a Hamiltonian for $\alpha$-particle. The isospin mixing of $\alpha$-particle immediately indicates that both of $^{12}$C and $^{16}$O receive the isospin mixing as well. 
It has also been shown that one can define such an effective isoscalar (IS) $E1$ operator that renormalizes  the isospin-mixing effect and in addition one can 
obtain the $E1$ matrix element  by evaluating 
 the matrix element of that effective IS $E1$ operator using the isospin-mixing free wave functions. 
This mechanism of introducing the isospin impurity has very recently been applied to predict the $E1$ decay width of the 9.64 MeV $3^-$ state of $^{12}$C~\cite{suzuki24}. The result turns out to be reasonable and encouraging. Accepting that the isospin impurity  can be treated in this way, one can focus on a standard spectroscopic study of the low-lying levels of $^{16}$O. 

Below $^{12}$C$+\alpha$ threshold, three positive-parity states and two negative-parity states exist~\cite{tilley93,database}; $0_1^+$ (0), \ $0_2^+$ (6.05), \  $2_1^+$ (6.92), and $3_1^-$ (6.13), \ $1_1^-$ (7.12), where the excitation energies in MeV are written in parentheses. In addition, two states lying above the threshold at 7.16 MeV,  $1_2^-$ (9.59) and $2^+_2$ (9.84)~\cite{tilley93,database}, may give  interference effects on the astrophysical $S$-factor. See, {\it e.g.} Refs.~\cite{angulo99,nacreII,deboer17}. There exist two direct processes, that is, the one-step transition from the $^{12}$C$+\alpha$ continuum to the ground state of $^{16}$O: One is the $E1$ transition from $P$-wave continuum and the other is the $E2$ transition from $D$-wave continuum. Other processes are all indirect or cascade transitions through some excited states of $^{16}$O.   It is conjectured that the direct processes dominate the cascade processes~\cite{pdesc87b,schurmann11,schurmann12,nacreII}. 

The direct processes mentioned above are not quite so simple because both of the $E1$ and $E2$ direct processes are considered to be strongly influenced by the subthreshold $1^-_1$ and $2^+_1$ states, respectively. See, {\it e.g.} Refs.~\cite{oulebsir12,nan24}. Because both 
subthreshold states are fairly strongly excited by $\alpha$-transfer reactions on $^{12}$C, they are considered to have large reduced-$\alpha$ widths.  Therefore their tails influence on the incident $\alpha-^{12}$C waves. The $1^-_1$ ($2^+_1$) state may play a role in the cascade transition as well. The 
strength of the $E1$ ($E2$) transition between the incident $S$-wave and the $1^-_1$ ($2_1^+$) state determines the reaction rate.  Another cascade transition via the $0_2^+$ state is discussed in Refs.~\cite{nacreII,matei06}. See also Refs.~\cite{dufour08,tischhauser02,shen20} for the $E2$ contributions. 

Extensive microscopic calculations of $^{12}$C$+\alpha$ 
two-cluster model were performed in Refs.~\cite{pdesc87,pdesc87b}, where $^{12}$C could be excited to the $2^+$ state described by the $j$-$j$ coupling shell-model and the isospin mixing was taken into account either by coupling to the nucleon channels of $^{15}$N$+p$ and $^{15}$O$+n$ or by introducing $\alpha$-cluster with $J^{\pi}=1^-, \, T=1$~\cite{pdesc86}. Prior to that calculation, $\alpha+^{12}$C$(0^+)$ and $\alpha+^{12}$C$(2^+)$ coupled-channel semi-microscopic  study~\cite{suzuki76a,suzuki76b,suzuki78} was carried out by the orthogonality-condition model~\cite{saito69}. Here, the wave function of $^{12}$C was assumed to be  an  $SU(3)$ (04) configuration. The energy spectra, electric transition probabilities, $\alpha$-decay widths and low-energy $\alpha+^{12}$C scattering observables were all reasonably well reproduced. However, that calculation included no isospin mixing. 

It appears very important and worthwhile to 
perform  $^{12}$C$(0^+)+\alpha$ and $^{12}$C$(2^+)+\alpha$ coupled-channel calculations with the inclusion of the isospin mixing as proposed in Ref.~\cite{suzuki21}.  Instead of simple shell-model wave functions for $^{12}$C~\cite{pdesc87,pdesc87b,suzuki76a,suzuki76b}, I intend to use microscopic $3\,\alpha$ cluster-model  wave functions  in which the motion of 3 $\alpha$-particles is not frozen to a particular configuration but can be determined 
dynamically.  For this purpose I need to extend the previous work~\cite{suzuki21} from a single projection scheme to the one including double projections of angular momenta. The angular momentum projection of a multi-cluster system in general demands a laborious task, and it is usually carried out numerically. See an example of  Ref.~\cite{kanada14}. I use correlated Gaussians (CGs) as basis functions because they are known to be 
versatile enough to represent different types of structure successfully and moreover allow analytic evaluations of various matrix elements~\cite{varga95,book98,book03,mitroy13}. I especially emphasize that double angular-momentum projections can be performed analytically including both the angular momentum of $^{12}$C and the  orbital angular momentum of  $^{12}$C$-\alpha$ relative motion. The purpose of this paper is to provide all the tools needed for a microscopic $^{12}$C$+\alpha$ coupled-channel study relevant to the $^{12}$C$(\alpha,\gamma)^{16}$O reaction. 

This paper is organized as follows. Section~\ref{formulation} defines  microscopic $^{12}$C$+\alpha$ basis functions described by CGs, introduces generator-coodinate-method kernels used in the microscopic $\alpha$-cluster model, and derives a transformation formula that makes it possible to represent the CG basis 
functions through the kernels. Section~\ref{impurity} recapitulates  Sec. 3 of Ref.~\cite{suzuki21} that introduces the isospin impurity of $\alpha$-particle and presents how to evaluate its $E1$ matrix element. 
Section~\ref{calculation.m.e.} shows some examples of how to  
calculate the CG matrix elements.   Section~\ref{summary} gives a summary and discussions. Appendix~\ref{negative.det} discusses a phase problem that may arise in transforming a Gaussian wave-packet (GWP) to the  CG.  All the formulas needed for the angular momentum projection are collected in Appendix~\ref{formulas}.

\section{Formulation}
\label{formulation}

\subsection{A microscopic $^{12}$C+$\alpha$ basis function}

Since any state of $^{16}$O of interest is assumed to be composed of $\alpha$-particles with $S=0$, its total spin 
is zero and consequently its total angular momentum is equal to the total orbital 
angular momentum $L$.  
The state $\Psi_{LM\pi}$ of $^{16}$O with the angular momentum $LM$ and the parity $\pi$ is described by a combination of  antisymmetrized $^{12}$C$+\alpha$ basis functions
\begin{align}
\Psi_{LM\pi}&=\sum_{l,u,A} c(l,u,A) \Psi_{l}^{LM\pi}(u,A),
\label{wf.total}\\
\Psi_{l}^{LM\pi}(u,A)&=\frac{1}{\sqrt{4!}^4}{\cal A}_{16}\Big\{ [f_{l_1}(u_1,A_1,{\bm \rho}_1)\times f_{l_2}(u_2,A_2,{\bm \rho}_2)]_{LM} \phi^{\rm in}(4\alpha)\Big\},
\label{basis.func}
\end{align}
where $l,u,A$ are variational parameters and stand for $(l_1,l_2),\, (u_1,u_2),\, (A_1, A_2)$, respectively. $l_1$ is the total angular momentum of $^{12}$C and $l_2$ the orbital angular momentum of the  $^{12}$C$-\alpha$ relative motion. $l_1+l_2$ is even or odd depending on $\pi=+$ or $-$. 
${\cal A}_{16}$ is the antisymmetrizer, $\frac{1}{\sqrt{16!}}\sum_p \epsilon(p)p$, summed over $16!$ 
permutations $p$ with their phases $\epsilon(p)$ and $\phi^{\rm in}(4\alpha)$ is the product of internal wave functions of 4 $\alpha$-particles
\begin{align}
\phi^{\rm in}(4\alpha)=\prod_{i=1}^4 \phi^{(0)}_{\alpha}(i).
\label{4alpha.intrinsic}
\end{align}
Here, $\phi^{(0)}_{\alpha}$ is a normalized, antisymmetrized $\alpha$-particle wave function free from its center-of-mass part. Its functional form is defined 
later in Eqs.~(\ref{Nalpha.WF}) and (\ref{wf.in.rel}) and also in Sec.~\ref{impurity}.   

The coordinates ${\bm \rho}_1$ and ${\bm \rho}_2$ stand for 
${\bm \rho}_1=
\left(\begin{array}{c}
{\bm x}_1  \\
{\bm x}_2 \\
\end{array}
\right)$ 
and ${\bm \rho}_2={\bm x}_3$, where ${\bm x}_i$'s are defined with the center-of-mass coordinates of $\alpha$-particles ${\bm R}_j$'s by 
\begin{align}
&\bm x_1=\bm R_1-\bm R_2,\ \ \  \bm x_2=\frac{1}{2}({\bm R}_1+{\bm R}_2)-{\bm R}_{3},\ \ \ \bm x_3=\frac{1}{3}({\bm R}_1+{\bm R}_2+{\bm R}_{3})-{\bm R}_4.
\label{rel.coord}
\end{align}
In  Eq.~(\ref{basis.func}), $f_{l_1}$ describes the motion of 3 $\alpha$-particles forming  $^{12}$C and $f_{l_2}$ the relative motion between $^{12}$C and a remaining $\alpha$-particle. 
$u_1$ is a 2-dimensional real vector and $A_1$ is a $2\times 2$ real, symmetric, positive-definite matrix, while $u_2$ is set to 1 and $A_2$ is a positive parameter. The functional form of $f_{l_i}$ is 
\begin{align}
f_{l_im_i}(u_i,A_i,{\bm \rho}_i)&={\cal Y}_{l_im_i}(\widetilde{u_i}{\bm \rho}_i)\,e^{-\frac{1}{2}\widetilde{{\bm \rho}_i}A_i{\bm \rho}_i},
\label{def.f-fun}
\\
{\cal Y}_{l_im_i}({\bm r})&=r^{l_i}Y_{l_im_i}(\hat{\bm r}),
\label{def.cal.Y}
\end{align}
where the tilde symbol $\,\widetilde{\,\,}\,$ indicates the transpose of a matrix or a column vector and $\hat{\bm r}=\frac{1}{r}{\bm r}$ stands for the polar and azimuthal angles of $\bm r$. The angular momenta of $l_1$ and $l_2$ are coupled to $LM$; 
\begin{align}
[f_{l_1}(u_1,A_1,{\bm \rho}_1)\times f_{l_2}(u_2,A_2,{\bm \rho}_2)]_{LM} =\sum_{m_1=-l_1}^{l_1}\sum_{m_2=-l_2}^{l_2}\langle l_1m_1 l_2 m_2|LM \rangle f_{l_1m_1}(u_1,A_1,{\bm \rho}_1) f_{l_2m_2}(u_2,A_2,{\bm \rho}_2).
\end{align} 

The basis function $\Psi_{l}^{LM\pi}(u,A)$ contains 7 parameters; 2 for $l=(l_1,l_2)$, 1 for $u_1$,  3 for $A_1$, and 1 for $A_2$. $u_1$ is normalized to $\widetilde{u_1}u_1=u_1(1)u_1(1)+u_1(2)u_1(2)=1$. The ratio between $u_1(1)$ and $u_1(2)$ 
controls the partial waves for $\alpha$-$\alpha$ motion and $2\alpha$-$\alpha$ motion in $^{12}$C, which is understood from the following identity (see Eq.~(6.108) of  Ref.~\cite{book98});
\begin{align}
{\cal Y}_{l_1m_1}(\widetilde{u_1}{\bm \rho}_1)&={\cal Y}_{l_1m_1}(u_1(1){\bm x}_1+u_1(2){\bm x}_2)\nonumber \\
&=\sum_{\lambda=0}^{l_1}\sqrt{\frac{4\pi(2l_1+1)!}{(2\lambda+1)!(2l_1-2\lambda+1)!}}(u_1(1))^{\lambda}(u_1(2))^{l_1-\lambda}[{\cal Y}_{\lambda}({\bm x}_1)\times {\cal Y}_{l_1-\lambda}({\bm x}_2)]_{l_1m_1}.
\label{Y.exp}
\end{align}
Though $u_2$ is fixed to 1, I keep its label for the sake of unifying the notation. The matrix $A_1$ primarily characterizes the size and shape of $^{12}$C, while $A_2$ controls the $^{12}$C$-\alpha$ distance.  The exponent 
$\widetilde{{\bm \rho}_i} A_i{\bm \rho}_i$ in Eq.~(\ref{def.f-fun}) stands for a scalar product, 
$\widetilde{{\bm \rho}_i}\! \cdot \!A_i{\bm \rho}_i$, i.e., $\widetilde{{\bm \rho}_1} A_1{\bm \rho}_1=A_1(1,1){\bm x}^{\,2}_1+2A_1(1,2){\bm x}_1\!\cdot\!{\bm x}_2+A_1(2,2){\bm x}^{\,2}_2$ and $\widetilde{{\bm \rho}_2} A_2{\bm \rho}_2=A_2{\bm x}^{\,2}_3$. Here, I point out another role of $A_1(1,2)$. 
If $A_1(1,2)$ is not zero, 
the term $e^{-A_1(1,2){\bm x_1}\cdot{\bm x_2}}$ modifies the partial-wave content between the $\alpha$-$\alpha$ motion and $2\alpha$-$\alpha$ motion mentioned above.  

It should be noted that $\Psi_{LM\pi}$ is flexible enough to include possible 
important sets of $(l_1l_2)$ for a given $L^{\pi}$: {\it e.g.} $(00)$ and $(22)$ for $0^+$, $(01)$ and $(21)$ for $1^-$, and $(02), (20)$, and $(22)$ for $2^+$. 

\subsection{A microscopic $4 \alpha$-cluster model}

A basic idea of evaluating the matrix element of an operator with $\Psi_l^{ LM\pi}(u,A)$ was proposed in Refs.~\cite{varga95,book98}. Its realization is lengthy, however, because of the antisymmetrizer. As shown in \cite{suzuki21}, the load is reasonably well 
reduced by expressing $\Psi_{l}^{LM\pi}(u,A)$ in terms of GWP used in a microscopic $N\alpha$-cluster model~\cite{brink66}. 

Let $\phi_{\bm S}^{\beta}$ denote a GWP centered at $\bm S$
\begin{align}
\phi_{\bm S}^{\beta}(\bm r)=\left(\frac{\beta}{\pi}\right)^{\frac{3}{4}}e^{-\frac{\beta}{2}(\bm r-\bm S)^2}.
\label{gwp}
\end{align}
By filling $\phi_{\bm S_i}^{\beta}(\bm r)$ with four spin-isospin states $\chi_j\ (j=1,...,4)$, I define a 4$\alpha$-particle cluster-model wave function~\cite{brink66,suzuki21}:
\begin{align}
\phi(\{\bm S\})={\cal A}_{16}\Big\{ 
\prod_{i=1}^4\Big( \phi_{\bm S_i}^{\beta} \chi_1 \phi_{\bm S_i}^{\beta} \chi_2  
\phi_{\bm S_i}^{\beta} \chi_3 \phi_{\bm S_i}^{\beta} \chi_4 \Big)\Big\}
=\frac{1}{{\sqrt{4!}}^4}{\cal A}_{16}\left\{
\prod_{i=1}^4 \phi_{\bm S_i}^{4\beta}(\bm R_i)\phi_{\alpha}^{(0)}(i)\right\}.
\label{Nalpha.WF}
\end{align}
Defining ${\bm s}_i$ $(i=1,2,3)$ from ${\bm S}_i$ $(i=1,...,4)$ in 
the same way as $\bm x_i$ is defined from $\bm R_i$ (see Eq.~(\ref{rel.coord})), I obtain 
\begin{align}
\prod_{i=1}^4 \phi_{\bm S_i}^{4\beta}(\bm R_i)\phi_{\alpha}^{(0)}(i)=\phi^{\rm in}(4\alpha)
\prod_{i=1}^4 \phi_{\bm S_i}^{4\beta}(\bm R_i)=
\phi^{\rm in}(4\alpha)\phi_{\overline{\bm S}}^{16\beta}(\bm R)\prod_{i=1}^{3}\phi_{\bm s_i}^{4\nu_i \beta}(\bm x_i),
\label{wf.in.rel}
\end{align}
where $\overline{\bm S}=\frac{1}{4}\sum_{i=1}^4\bm S_i$ and $\nu_i=\frac{i}{i+1}$.  The center-of-mass motion is factored out:  
\begin{align}
\phi(\{\bm S\})=\phi^{\rm in}(\{\bm s\})\phi^{16\beta}_{\overline{\bm S}}(\bm R).
\label{intri.cm.sepa}
\end{align}

By letting $\bm x$ and $\bm s$ stand for the set $\{\bm x_1,\bm x_2,\bm x_3\}$ and $\{\bm s_1,\bm s_2,\bm s_3\}$, respectively, the product of the GWPs for the 4 $\alpha$-particle relative motion reads as
\begin{align}
\prod_{i=1}^{3}\phi_{\bm s_i}^{4\nu_i \beta}(\bm x_i)&=\Big({\rm det}(\pi^{-1}\Gamma) \Big)^{\frac{3}{4}}\, e^{-\frac{1}{2}\tilde{\bm x}\Gamma \bm x+\tilde{\bm x}\Gamma \bm s-\frac{1}{2}\tilde{\bm s}\Gamma \bm s},
\label{prod.GWP}
\end{align}
where $\Gamma$ is a $3\times 3$ diagonal matrix,
\begin{align}
\Gamma&=(4 \beta \nu_i\delta_{i,j}),\\
{\rm det}\, \Gamma&=\frac{1}{4}(4\beta)^{3}.
\end{align}
By removing the total c.m. function $\phi_{\overline{\bm S}}^{16\beta}(\bm R)$ 
from $\phi(\{\bm S\})$, 
the intrinsic wave function of $4\alpha$-particle system takes the form 
\begin{align}
\phi^{\rm in}(\{\bm s\}) = \Big({\rm det}(\pi^{-1}\Gamma) \Big)^{\frac{3}{4}} 
 \frac{1}{{\sqrt{4!}}^4} 
{\cal A}_{16}\left\{ e^{-\frac{1}{2}\tilde{\bm x}\Gamma \bm x+\tilde{\bm x}\Gamma \bm s-\frac{1}{2}\tilde{\bm s}\Gamma \bm s}\phi^{\rm in}(4\alpha) \right\}.
\label{gcm.wf.nalpha}
\end{align}
Note that $\bm s$ is a set of parameter variables independent of  the permutations.

\subsection{Generating $^{12}$C$+\alpha$ basis function from $4\alpha$-cluster model basis}

I express $\Psi_{l}^{LM\pi}(u,A)$ in terms of  $\phi^{\rm in}(\{\bm s\})$. As shown below, the function of type~\cite{varga95,book98} 
\begin{align}
g(\alpha_i \bm e_i u_i, A_i, {\bm \rho}_i)=e^{-\frac{1}{2}\widetilde{{\bm \rho}_i}A_i{\bm \rho}_i+\alpha_i \bm e_i \widetilde{u_i}{\bm \rho}_i}
\label{def.g-func}
\end{align}
plays a vital role to generate $f_{l_im_i}$ as well as to bridge the cluster-model 
basis. Here, $\bm e_i$ is a 3-dimensional unit vector, ${\bm e_i}\cdot{\bm e}_i=1$, and $\alpha_i \bm e_i \widetilde{u_i}{\bm \rho}_i$ stands for a scalar product, 
$\alpha_1 \bm e_1 \widetilde{u_1}{\bm \rho}_1=\alpha_1 \bm e_1 \cdot (u_1(1){\bm x}_1+u_1(2){\bm x}_2)$ and $\alpha_2 \bm e_2 \widetilde{u_2}{\bm \rho}_2=\alpha_2 \bm e_2 \cdot u_2{\bm x}_3$. I begin by expressing the basis function~(\ref{basis.func}) in terms of $g$. A useful relation is~\cite{varga95,book98,suzuki08} 
\begin{align}
\int d{\bm e}\, Y_{lm}(\bm e){\cal D}^l(\alpha) \, e^{\alpha \bm e\cdot \bm r}=\frac{1}{B_l}{\cal Y}_{lm}(\bm r),
\end{align}
where $\int d{\bm e}$ stands for the integration over the total solid angle and 
\begin{align}
{\cal D}^{l}(\alpha)=\frac{1}{l!}\frac{\partial^{l}}{\partial\alpha^{l}}\Big|_{\alpha=0},\ \ \ \ \ B_{l}=\frac{(2l+1)!!}{4\pi}.
\end{align}
The function $f_{l_im_i}$ of Eq.~(\ref{def.f-fun}) is generated from $g$ by  
\begin{align}
f_{l_im_i}(u_i,A_i,{\bm \rho}_i)= B_{l_i}\int d\bm e_i \, Y_{l_im_i}(\bm e_i)\, {\cal D}^{l_i}(\alpha_i)\, g(\alpha_i \bm e_i u_i, A_i, {\bm \rho}_i).
\label{gtof}
\end{align}

I define a 3-dimensional column vector, $\alpha {\bm e}u$, and a 
$3\times 3$ positive-definite symmetric matrix, $A$, by 
\begin{align}
\alpha {\bm e} u= \left(\begin{array}{c}
\alpha_1 {\bm e}_1 u_1\\
\alpha_2 {\bm e}_2 u_2 \\
\end{array}
\right)= \left(\begin{array}{c}
\alpha_1 {\bm e}_1 u_1(1)\\
\alpha_1 {\bm e}_1 u_1(2)\\
\alpha_2 {\bm e}_2 u_2 \\
\end{array}
\right),
\ \ \ \ \ 
A=\left(\begin{array}{cc}
A_1 & 0 \\
0 & A_2 \\
\end{array}
\right)=\left(\begin{array}{ccc}
A_1(1,1) & A_1(1,2) & 0 \\
A_1(1,2) & A_1(2,2) & 0 \\
0 & 0 & A_2 \\
\end{array}
\right).
\label{uAdef}
\end{align} 
The use of $\prod_{i=1}^2 
g(\alpha_i \bm e_i u_i, A_i, {\bm \rho}_i)=g(\alpha \bm e u, A, \bm x)$ with 
${\bm x} = \Big(\begin{array}{c}
{\bm \rho}_1\\
{\bm \rho}_2 \\
\end{array}
\Big)=\Bigg(\begin{array}{c}
{\bm x}_1\\
{\bm x}_2 \\
{\bm x}_3 \\
\end{array}
\Bigg)$ and Eq.~(\ref{gtof}) leads to 
\begin{align}
[f_{l_1}(u_1,A_1,{\bm \rho}_1)\!\times \!f_{l_2}(u_2,A_2,{\bm \rho}_2)]_{LM}
=B_{l_1}B_{l_2}\iint d{\bm e}_1 d{\bm e}_2 \, 
[Y_{l_1}({\bm e}_1)\!\times \! Y_{l_2}({\bm e}_2)]_{LM} 
\prod_{i=1}^2 {\cal D}^{l_i}(\alpha_i)\, g(\alpha \bm e u, A, \bm x).
\label{ff.to.g}
\end{align}
The basis function~(\ref{basis.func}) is therefore generated through $g(\alpha \bm e u, A, \bm x)$ as follows;
\begin{align}
\Psi_{l}^{LM\pi}(u,A)=B_{l_1}B_{l_2}\iint d{\bm e}_1 d{\bm e}_2 \, 
[Y_{l_1}({\bm e}_1)\!\times \! Y_{l_2}({\bm e}_2)]_{LM} 
\prod_{i=1}^2 {\cal D}^{l_i}(\alpha_i)\,
\frac{1}{\sqrt{4!}^4}{\cal A}_{16}\Big\{ g(\alpha \bm e u, A, \bm x)\, \phi^{\rm in}(4\alpha)\Big\}.
\label{first-step}
\end{align}
Specifying the partial waves $(l_1,l_2)$ and coupling them to the total angular momentum $LM$ are carried out by such algebraic operations with respect to $\alpha_i$'s and ${\bm e}_i$'s that are independent of the permutations. 

Next I express  $\frac{1}{\sqrt{4!}^4}{\cal A}_{16}\Big\{ g(\alpha \bm e u, A, \bm x)\, \phi^{\rm in}(4\alpha)\Big\}$ in terms of the integral transform of $\phi^{\rm in}(\{ {\bm s} \})$ of Eq.~(\ref{gcm.wf.nalpha}).   As shown in \cite{suzuki21}, the integral of $e^{ -\frac{1}{2}\tilde{\bm x}\Gamma {\bm x} + \widetilde{\bm x}\Gamma {\bm s} -\frac{1}{2} \widetilde{\bm s}\Gamma {\bm s}}$ times $Cg(\bm v, Q, \bm s)$ with respect to $\bm s$ can be made equal to $g(\alpha \bm e u, A, \bm x)$: 
\begin{align}
g(\alpha \bm e u, A, \bm x) & = \Big({\rm det}({\pi}^{-1}\Gamma )\Big)^{\frac{3}{4}}\, C \int d{\bm s} \, g(\bm v, Q, \bm s) \, 
e^{-\frac{1}{2}\widetilde{\bm x}\Gamma {\bm x} 
+ \widetilde{\bm x}\Gamma {\bm s} -\frac{1}{2} \widetilde{\bm s}\Gamma {\bm s}} 
\nonumber \\
& = \Big({\rm det}({\pi}^{-1}\Gamma )\Big)^{\frac{3}{4}}\, C \Big(\frac{(2\pi)^3}{{\rm det}(Q+\Gamma)}\Big)^{\frac{3}{2}}e^{-\frac{1}{2}\widetilde{\bm x}\Gamma {\bm x}
+\frac{1}{2}\widetilde{(\bm v +\Gamma {\bm x})} (Q+\Gamma)^{-1}( {\bm v +\Gamma {\bm x}})}.
\label{xtos}
\end{align}
Here, $\int d{\bm s}$ stands for $\int \!\int \!\int d{\bm s}_1 d{\bm s}_2 d{\bm s}_3$.  The unknowns, $Q, {\bm v}$, and $C$, are found to be 
\begin{align}
Q&=\Gamma (\Gamma -A)^{-1}\Gamma-\Gamma=A(\Gamma-A)^{-1}\Gamma,\nonumber \\
{\bm v}&=\Gamma (\Gamma -A)^{-1}\alpha {\bm e}u=\alpha {\bm e} \Gamma (\Gamma -A)^{-1} u,\nonumber \\
C&=\Big(\frac{{\rm det}\Gamma}{4\pi}\Big)^{\frac{9}{4}} [{\rm det}\, (\Gamma- A)]^{-\frac{3}{2}}\, e^{-\frac{1}{2}\widetilde{\alpha {\bm e}u}(\Gamma -A)^{-1} \alpha {\bm e}u}.
\label{def.Qv}
\end{align} 
Since $\Gamma$ is diagonal and $A$ has the form of Eq.~(\ref{uAdef}), both $\Gamma$ and $(\Gamma -A)^{-1}$ have the same form as $A$, 
\begin{align}
\Gamma=\left(\begin{array}{cc}
\Gamma_1 & 0 \\
0 & \Gamma_2 \\
\end{array}
\right),
\ \ \ \ \ \ \ 
(\Gamma -A)^{-1} =\left(\begin{array}{cc}
(\Gamma_1-A_1)^{-1} & 0 \\
0 & (\Gamma_2-A_2)^{-1} \\
\end{array}
\right).
\end{align}
Note that the matrix $Q$ is symmetric. The $\alpha$-dependence of $C$ reads as 
\begin{align} 
C=C_0 \, e^{-\frac{1}{2}\sum_{i=1}^2\widetilde{u_i}(\Gamma_i-A_i)^{-1} u_i \, \alpha_i^{\,2} },\ \ \ \ \ \ \ C_0=\Big(\frac{{\rm det}\Gamma}{4\pi}\Big)^{\frac{9}{4}}\Big({\rm det}(\Gamma-A)\Big)^{-\frac{3}{2}}.
\end{align}
To conclude, $\Psi_{l}^{LM\pi}(u,A)$ is generated by the integro-differential operation on the integral transform of $\phi^{\rm in}(\{ {\bm s} \})$;
\begin{align}
\Psi_{l}^{LM\pi}(u,A)&=  B_{l_1}B_{l_2}  C_0 \iint d{\bm e}_1 d{\bm e}_2 
[Y_{l_1}({\bm e}_1)\!\times \! Y_{l_2}({\bm e}_2)]_{LM} 
\prod_{i=1}^2 {\cal D}^{l_i}(\alpha_i) \nonumber \\
&\ \ \times e^{-\frac{1}{2}\sum_{i=1}^2\widetilde{u_i}(\Gamma_i-A_i)^{-1}u_i \, \alpha_i^{\,2}} \int d{\bm s} \,g(\bm v, Q, \bm s)\phi^{\rm in}(\{ {\bm s} \}).
\label{2ndstep}
\end{align}
Note that the 3-dimensional column vector ${\bm v}$ depends on $\alpha_i{\bm e}_i\, (i=1,2)$. It may happen that ${\rm det}(\Gamma-A)$ of Eq.~(\ref{def.Qv}) is negative. 
This problem is taken up later when a final formula to calculate the matrix element is given in Eqs.~(\ref{CG.me.final}), (\ref{def.cal.G}), and (\ref{def.cal.U}), and will be discussed in Appendix~\ref{negative.det}.

\section{Electric dipole transitions}
\label{impurity}

The state $\Psi_{LM\pi}$ of Eq. (\ref{wf.total}) has so far been assumed to have isospin $T=0$, so that it has no isovector (IV) $E1$ matrix 
elements. It is crucially important to take up this problem in order to evaluate 
due contribution of the $E1$ transition to $^{12}$C$+\alpha$ radiative-capture cross section. Motivated by this problem, 
I proposed in Ref.~\cite{suzuki21} a  way of introducing isospin $T=1$ components into $\Psi_{LM\pi}$ through an extension of the wave function of $\alpha$-particle.  Moreover, its application has very recently been made to predict  the $E1$ decay width of the 9.64 MeV $3^-$ state of $^{12}$C~\cite{suzuki24}.  Though the magnitude of the $T=1$ component is small, the effect of the isospin mixing on the $E1$ transition  is so large as to account for more than half of the $E1$ decay width of the $3^-$ state. As a result, it turns out that the calculated $E1$ strength including both IV and IS nature relative to the Weisskopf unit is close to the corresponding ratios of the $E1$ transitions of the low-lying $1^-$ states in $^{16}$O. In what follows I recapitulate the basic idea of introducing the isospin mixing into $\Psi_{LM\pi}$.

Suppose that the wave function of $\alpha$-particle is a combination of both $T=0$ and 1 components~\cite{suzuki21}:
\begin{align}
\phi_{\alpha}=\sqrt{1-\epsilon^2}\phi^{(0)}_{\alpha} + \epsilon \phi^{(1)}_{\alpha}.
\label{def.epsilon}
\end{align}
Here, $\phi^{(0)}_{\alpha}=\phi_{\alpha}^{\rm orb}\Omega$ is the major part described by the $(0s)^4$ configuration with its center-of-mass motion being removed:
\begin{align}
\phi_{\alpha}^{\rm orb}&=\big(\frac{\beta^3}{4\pi^3}\big)^{\frac{3}{4}}e^{-\frac{\beta}{2}\sum_{i=1}^4(\bm r_i-\bm R_1)^2},\\
\Omega&=\frac{1}{\sqrt{2}}\{|(00)00,(11)00\rangle -|(11)00,(00)00\rangle \},
\end{align}
where ${\bm R_1}=\frac{1}{4}\sum_{i=1}^4\bm r_i$ is the center-of-mass coordinate of $\alpha$-particle and $\Omega$ is the totally antisymmetric spin-isospin function with $S=0,\, T=0$. Here, $\Omega$ is expressed in the 
representation, $|(S_{12}S_{34})SM_S, (T_{12}T_{34})TM_T\rangle$, where {\it e.g.} $S_{12}$ stands for the spin resulting from coupling  
the spins of the first and second nucleons, and the spins of $S_{12}$ and $S_{34}$ are coupled to $SM_S$. 
In its simplest version $\phi^{(1)}_{\alpha}$ 
has 2$\hbar \omega$ excitation built on $\phi_{\alpha}^{\rm orb}$~\cite{suzuki21}: 
\begin{align}
&\phi^{(1)}_{\alpha}=\frac{\beta}{3\sqrt{2}}\phi^{\rm orb}_{\alpha}
\sum_{i=1}^4 (\bm R_1 -\bm r_i)^2\Omega_i,
\end{align}
where $\Omega_i$'s all have $S=0, \, T=1$ and they are defined by 
\begin{align}
\Omega_1&=-|(00)00,(01)10\rangle +\sqrt{2}|(00)00,(11)10 \rangle -\sqrt{3}|(11)00,(10)10\rangle,\nonumber \\
\Omega_2&=\ \ \,|(00)00,(01)10\rangle +\sqrt{2}|(00)00,(11)10\rangle +\sqrt{3}|(11)00,(10)10\rangle,\nonumber \\
\Omega_3&=-|(00)00,(10)10\rangle -\sqrt{2}|(00)00,(11)10\rangle -\sqrt{3}|(11)00,(01)10\rangle,\nonumber \\
\Omega_4&=\ \ \,|(00)00,(10)10\rangle -\sqrt{2}|(00)00,(11)10\rangle +\sqrt{3}|(11)00,(01)10\rangle.
\end{align}
Note that $\sum_{i=1}^4 \Omega_i=0$ and $\langle \Omega_i|\Omega_j\rangle=8\delta_{i,j}-2$. An important property of $\Omega_i$  is 
\begin{align}
\langle \Omega|\frac{1}{2}-{t_j}_{3}|\Omega_i\rangle=\frac{1}{\sqrt{6}}(1-4\delta_{i,j}),
\label{me.t3}
\end{align}
where ${t_j}_{3}$ is the third component of the $j$-th nucleon's isospin.  Equation~(\ref{me.t3}) plays a key role 
in accounting for the isospin mixing.
The isospin-mixing coefficient $\epsilon$ in Eq.~(\ref{def.epsilon})  was estimated by diagonalyzing a Hamitonian for $\alpha$-particle with the two states of $\phi^{(0)}_{\alpha}$ and $\phi^{(1)}_{\alpha}$~\cite{suzuki21}. 
The value of $\epsilon$ turns out to be about $-4.2\times 10^{-3}$, that is, the probability of the $T=1$ component in $\alpha$-particle is at most on the order of 2$\times 10^{-3}$\%. 

Because $\epsilon$ is sufficiently small, $\Psi_{LM\pi}$ of Eq.~(\ref{wf.total}) is defined using $\phi^{(0)}_{\alpha}$ as indicated
in Eq.~(\ref{4alpha.intrinsic}), and all the matrix elements but the 
IV $E1$ matrix element can safely be obtained by neglecting the isospin impurity of $\alpha$-particles. The isospin mixing in $^{16}$O is introduced by replacing one of 
$\phi^{(0)}_{\alpha}$'s with $\phi^{(1)}_{\alpha}$:   
\begin{align}
\phi^{\rm in}(4\alpha)\approx \prod_{i=1}^4 \phi^{(0)}_{\alpha}(i)+
\epsilon \sum_{i=1}^4 \phi^{(1)}_{\alpha}(i) \prod_{j\neq i}^4\phi^{(0)}_{\alpha}(j).
\end{align}
Therefore, a state ${\Psi'}_{LM\pi}$  that includes the $T=1$ mixing is constructed from $\Psi_{LM\pi}$ to a very good approximation by 
\begin{align}
 {\Psi'}_{LM\pi}\approx \Psi_{LM\pi}+\epsilon \sum_{i=1}^4 \Psi_{LM\pi}(i),
\end{align}
where $\Psi_{LM\pi}(i)$ is defined by replacing $\phi^{(0)}_{\alpha}(i)$ with 
$\phi^{(1)}_{\alpha}(i)$ in Eq.~(\ref{4alpha.intrinsic}), whereas the rest of the $\alpha$-particle wave functions remain unchanged. In this way $\Psi_{LM\pi}$ is extended to include the isospin $T=1$ component. The $E1$ matrix element reads as
\begin{align}
&\langle {\Psi'}_{L'M'\pi'} |E_{1\mu}({\rm IV})+E_{1\mu}({\rm IS})|{\Psi'}_{LM\pi}\rangle \nonumber \\
&\approx \langle \Psi_{L'M'\pi'}|E_{1\mu}({\rm IS})|\Psi_{LM\pi}\rangle
+\epsilon \sum_{i=1}^4\Big\{ \langle \Psi_{L'M'\pi'}|E_{1\mu}({\rm IV})|\Psi_{LM\pi}(i)\rangle +\langle \Psi_{L'M'\pi'}(i)|E_{1\mu}({\rm IV})|\Psi_{LM\pi}\rangle \Big\},
\end{align}
where $E_{1\mu}({\rm IS})$ is the IS $E1$ operator next to the  IV $E1$ operator, $E_{1\mu}({\rm IV})$, of leading order. See Ref.~\cite{baye12}. The second term on the right-hand side is the IV $E1$ matrix element that appears due to the isospin mixing. As shown in Ref.~\cite{suzuki21}, the term is converted, owing to Eq.~(\ref{me.t3}), to a simple matrix element as follows;
\begin{align}
\sum_{i=1}^4  \Big\{\langle \Psi_{L'M'\pi'} | E_{1\mu}({\rm IV})|\Psi_{LM\pi}(i) \rangle
+\langle \Psi_{L'M'\pi'}(i)|E_{1\mu}({\rm IV})|\Psi_{LM\pi}\rangle \Big\}
=2 \langle \Psi_{L'M'\pi'}| E_{1\mu}^{\rm eff}({\rm IS})|\Psi_{LM\pi} \rangle,
\end{align}
where $E_{1\mu}^{\rm eff}({\rm IS})$ is an effective IS $E1$ operator that renormalizes the isospin-impurity components; 
\begin{align}
E_{1\mu}^{\rm eff}({\rm IS})=e\frac{2\beta}{3\sqrt{3}}\sum_{i=1}^4 \sum_{j=4i-3}^{4i}({\bm R}_i-{\bm r}_j)^2{\cal Y}_{1\mu}({\bm R}_i-{\bm r}_j).
\end{align}
Here, ${\bm R}_i=\frac{1}{4}\sum_{j=4i-3}^{4i}{\bm r}_j$ is the center-of-mass coordinate of the $i$-th $\alpha$-particle.

The $E1$ operator consists of a few terms up to the first order 
beyond the long wavelength approximation~\cite{baye12}. A comparative study of relative importance of those terms including $E_{1\mu}^{\rm eff}({\rm IS})$ has been made for the transition of the 9.64 MeV $3^-$ state to 
the 4.44 MeV 2$^+$ state of $^{12}$C~\cite{suzuki24}. What is natural and appealing of the study is that $\epsilon$ is determined independent 
of the $E1$ transition data of $^{12}$C.  It is now possible to take into account both the IV and IS $E1$ operators to see the extent to which they 
contribute to the $^{12}$C$(\alpha,\gamma)^{16}$O radiative-capture reaction.  
 This is an advantage compared to 
the approach using only the $E_{1\mu}({\rm IV})$ operator~\cite{pdesc87}.

\section{Calculation of $^{12}$C$+\alpha$ matrix elements}
\label{calculation.m.e.}

The matrix element of an operator ${\cal O}_{\kappa \mu}$ between the basis functions reads as 
\begin{align}
&\langle \Psi^{L'M'\pi'}_{l'}(u',A') |{\cal O}_{\kappa \mu}| \Psi_{l}^{LM\pi}(u,A)\rangle\nonumber \\
&= {\cal F} \int \cdots \int d{\bm e}_1 \cdots d{\bm e}_4 \, {\cal D}_{l_1\cdots l_4}^{LML'M'}(\alpha_1 \cdots \alpha_4)\,
 e^{-\frac{1}{2}\sum_{i=1}^4\widetilde{u_i}(\Gamma_i-A_i)^{-1} u_i \, \alpha_i^{\,2}}\, I(\alpha_1{\bm e}_1,\cdots,\alpha_4{\bm e}_4),
\label{CG.me.formula}
\end{align}
where $\Psi^{\pi'L'M'}_{l'}(u',A')$ is defined in the same way as $\Psi_{l}^{LM\pi}(u,A)$ of Eq.~(\ref{2ndstep}) by either putting $'$ and/or changing labels $1,2$ to $3,4$. Note also that 
$\Gamma_3$ and $\Gamma_4$ are introduced to stand for $\Gamma_1$ and $\Gamma_2$, respectively.  The following  abbreviated notations are introduced for the sake of simplicify:
\begin{align}
{\cal F}&=\big(\prod_{i=1}^4 B_{l_i}\big) C_0 C_0', \nonumber \\
\int \cdots \int d{\bm e}_1 \cdots d{\bm e}_4 \,  {\cal D}_{l_1\cdots l_4}^{LML'M'}(\alpha_1 \cdots \alpha_4)
&= \int \cdots \int d{\bm e}_1 \cdots d{\bm e}_4 \,[Y_{l_3}({\bm e}_3)\!\times \! Y_{l_4}({\bm e}_4)]^*_{L'M'} [Y_{l_1}({\bm e}_1)\!\times \! Y_{l_2}({\bm e}_2)]_{LM} \prod_{i=1}^4 {\cal D}^{l_i}(\alpha_i),\nonumber \\
I(\alpha_1{\bm e}_1,\cdots,\alpha_4{\bm e}_4) &= \iint d{\bm s} \,d{\bm s}' \,g({\bm v}', Q', {\bm s}') g(\bm v, Q, \bm s)\langle \phi^{\rm in}(\{ {\bm s}' \})|{\cal O}_{\kappa \mu}|\phi^{\rm in}(\{ {\bm s} \}) \rangle.
\label{abbr.notation}
\end{align}
Note that $\langle \phi^{\rm in}(\{ \bm s' \})| {\cal O}_{\kappa \mu}|\phi^{\rm in}(\{ {\bm s}\}) \rangle$ in Eq.~(\ref{abbr.notation}) is nothing but the generator-coordinate-method kernel for ${\cal O}_{\kappa \mu}$, which can easily be obtained as a function of $\bm s$ and $\bm s'$ \cite{suzuki21}. 
Equation~(\ref{CG.me.formula}) confirms that the desired matrix element can be 
obtained by the following operations; (I) the integration 
with respect to $\bm s$ and ${\bm s}'$,  (II) the differentiation with respect to $\alpha_1,...,\alpha_4$, and (III) the integration with respect to ${\bm e}_1, ..., {\bm e}_4$. I unify some notations to make the $\alpha, \bm e$-dependence 
of $I(\alpha_1{\bm e}_1,\cdots,\alpha_4{\bm e}_4)$ explicit, which is useful 
for achieving the operations systematically. 

For the step (I), let $\bm t_i\, (i=1,2,\ldots,6)$ stand for ${\bm s}_1, {\bm s}_2, {\bm s}_3, {\bm s}'_1, {\bm s}'_2, {\bm s}'_3$. The kernel 
$\langle \phi^{\rm in}(\{ \bm s' \})| {\cal O}_{\kappa \mu}|\phi^{\rm in}(\{ {\bm s} \}) \rangle$ of interest consists of a number of terms of the form \cite{suzuki21}
\begin{align}
P({\bm t})\, e^{-\frac{1}{2}{\tilde {\bm t}}W {\bm t}},\ \ \ \ \ \ \ W=\left(\begin{array}{cccc}
W_{11} & W_{12} & W_{13} & W_{14} \\
W_{21} & W_{22} & W_{23} & W_{24} \\
W_{31} & W_{32} & W_{33} & W_{34} \\
W_{41} & W_{42} & W_{43} & W_{44} \\ 
\end{array}
\right),
\label{kernel.form}
\end{align}
where $P({\bm t})$ is a polynomial of $\bm t$ and $W$ is a $6\times 6$ symmetric matrix: the size of the submatrix $W_{ij}$ is $d(i)\!\times\! d(j)$ with $d(k)\!=\!2(\delta_{k,1}\!+\!\delta_{k,3})+\delta_{k,2}+\delta_{k,4}$. The product 
$g(\bm v', Q', \bm s')g({\bm v}, Q, {\bm s})$ of Eq.~(\ref{abbr.notation})
is equal to $e^{-\frac{1}{2}\tilde{\bm t}{\cal Q}{\bm t}+\widetilde{{\bm \upsilon}}\bm t}$, where  $\bm {\upsilon }$ is a 6-dimensional vector and  ${\cal Q}$ is a $6\times 6 $ symmetric matrix that are respectively defined by 
\begin{align}
&{\bm {\upsilon}}=
\left(\begin{array}{c}
{\bm v} \\
{\bm v}' \\
\end{array}
\right)=
\left(\begin{array}{c}
\alpha_1 {\bm e}_1 \Gamma_1 (\Gamma_1-A_1)^{-1} u_1 \\
\alpha_2 {\bm e}_2 \Gamma_2 (\Gamma_2-A_2)^{-1} u_2 \\
\alpha_3 {\bm e}_3 \Gamma_3 (\Gamma_3-A_3)^{-1} u_3 \\
\alpha_4 {\bm e}_4 \Gamma_4 (\Gamma_4-A_4)^{-1} u_4 \\
\end{array}
\right),\nonumber \\
&{\cal Q}=
\left(\begin{array}{cc}
Q & 0  \\
0 & Q' \\
\end{array}
\right)=
\left(\begin{array}{cccc}
A_1(\Gamma_1-A_1)^{-1}\Gamma_1 & 0 & 0 & 0 \\
0 & A_2(\Gamma_2-A_2)^{-1}\Gamma_2 & 0 & 0 \\
0 & 0 & A_3(\Gamma_3-A_3)^{-1}\Gamma_3 & 0 \\
0 & 0 & 0 & A_4(\Gamma_4-A_4)^{-1}\Gamma_4 \\ 
\end{array}
\right).
\end{align}
The integration with respect to $\bm s$ and $\bm s'$ in 
Eq.~(\ref{abbr.notation}) is compactly expressed as
\begin{align}
I(\alpha_1{\bm e}_1,\cdots,\alpha_4{\bm e}_4) = \int d{\bm t}  \, e^{-\frac{1}{2}\tilde{\bm t}{\cal Q}{\bm t}+\widetilde{{\bm \upsilon}}\bm t}P(\bm t)e^{-\frac{1}{2}\tilde{\bm t}W{\bm t}}.
\label{t-integral}
\end{align}
Table 2 of Ref.~\cite{suzuki21} lists $I(\alpha_1{\bm e}_1,\cdots,\alpha_4{\bm e}_4)$ for $P(\bm t)$ of practical importance. $I(\alpha_1{\bm e}_1,\cdots,\alpha_4{\bm e}_4)$ turns out to be given by a combination of  five types of functions:  
\begin{align}
\Big(\frac{(2\pi)^6}{{\rm det}Z}\Big)^{\frac{3}{2}}e^{\frac{1}{2}\widetilde{{\bm \upsilon}}Z^{-1}{\bm \upsilon}}F_i(\bm \upsilon),
\end{align}
where $Z=W+{\cal Q}$ is a symmetric matrix and  
\begin{align}
&F_1(\bm \upsilon)=1,\ \ \ \ \ \ \ F_2(\bm \upsilon)=\widetilde{\bm \upsilon}X{\bm \upsilon},\ \ \ \ \ \ \ F_3(\bm \upsilon)={\cal Y}_{\kappa \mu}(\widetilde{\omega}\bm \upsilon),\nonumber \\
&F_4(\bm \upsilon)=\widetilde{\bm \upsilon}X{\bm \upsilon}{\cal Y}_{\kappa \mu}(\widetilde{\omega}\bm \upsilon),\ \ \ \ \ \ \ F_5(\bm \upsilon)=(\widetilde{\bm \upsilon}X{\bm \upsilon})(\widetilde{\bm \upsilon}Y{\bm \upsilon}){\cal Y}_{\kappa \mu}(\widetilde{\omega}\bm \upsilon).
\end{align}
Here, $X$ and $Y$ are $6\times 6$ symmteric matrices that have the same structure as $W$ of Eq.~(\ref{kernel.form}) and $\omega$ is a 6-dimensional column vector.  To conclude, $I(\alpha_1{\bm e}_1,\cdots,\alpha_4{\bm e}_4)$ is characterized by two types of $\alpha, {\bm e}$-dependent terms: one is  $\widetilde{\omega}\bm \upsilon$ of linear type, and the other is $\widetilde{\bm \upsilon}Z^{-1}{\bm \upsilon}$ of quadratic type. Their $\alpha, {\bm e}$-dependence is 
respectively expressed by $\sum_{i=1}^4 \gamma_i\alpha_i{\bm e}_i$ and $\sum_{1\le i< j\le 4} \rho_{ij}\alpha_i \alpha_j {\bm e}_i\cdot{\bm e}_j$, where $\gamma_i$ and $\rho_{ij}$ are appropriate coefficients. 

The operation of step (II), 
$\int \cdots \int d{\bm e}_1 \cdots d{\bm e}_4 \, {\cal D}_{l_1\cdots l_4}^{LML'M'}(\alpha_1 \cdots \alpha_4)$,  
on $e^{-\frac{1}{2}\sum_{i=1}^4\widetilde{u_i}(\Gamma_i-A_i)^{-1} u_i \, \alpha_i^{\,2}}I(\alpha_1{\bm e}_1,\cdots,\alpha_4{\bm e}_4)$ is carried out by expanding 
$I(\alpha_1{\bm e}_1,\cdots,\alpha_4{\bm e}_4)$ in the powers of $\alpha_i{\bm e}_i\, (i=1,\cdots,4)$. Because ${\cal D}^{l_i}(\alpha_i)$ gives non-vanishing contributions only when it acts on terms containing $\alpha_i^{l_i}$, and because $\alpha_i$ and ${\bm e}_i$ are connected together in the generating function~(\ref{def.g-func}), those terms with the $l_i$-th power of $\alpha_i$ should contain  at most $l_i$ ${\bm e}_i$'s. Since the maximum possible angular momentum allowed by at most $l_i$ ${\bm e}_i$'s is $l_i$, the angle integration, $\int d{\bm e}_i Y_{l_i}({\bm e}_i)$, of step (III) gives non-vanishing contributions only when $l_i$ ${\bm e}_i$'s are successively coupled by the so-called stretched or aligned coupling
\begin{align}
[Y_a({\bm e}_i)\times Y_b({\bm e}_i)]_{lm} &\Rightarrow C(a\, b;l)Y_{lm}({\bm e}_i)\delta_{l,a+b},\nonumber \\
C(a\, b;c)&=\sqrt{\frac{(2a+1)(2b+1)}{4\pi (2c+1)}}\langle a\, 0\, b\, 0|c\,0\rangle.
\label{stretched.coupling}
\end{align}
By the symbol $\Rightarrow$ I mean both sides are equal as long as the stretched coupling is valid. It is also useful to note that the power of 
$({\bm e}_i\cdot {\bm e}_j)$ reduces to  (see Eq. (6.18) of Ref.~\cite{book98})
\begin{align}
({\bm e}_i\cdot {\bm e}_j)^n \Rightarrow \frac{n!}{B_n} (-1)^n\sqrt{2n+1}[Y_n({\bm e}_i)\times Y_n({\bm e}_j)]_{00},
\label{YY.power}
\end{align} 
which is also proved by induction with the help of a special 
Clebsch-Gordan coefficient for  integers $a$ and $b$, 
\begin{align}
\langle a\, 0\, b\, 0|a\!+\!b \, 0\rangle
=\sqrt{\frac{(2a)!(2b)!}{(2(a\!+\!b))!}}\frac{{(a\!+\!b)!}} {a!b!}.
\end{align}
Since $e^{-\frac{1}{2}\sum_{i=1}^4\widetilde{u_i}(\Gamma_i-A_i)^{-1} u_i \, \alpha_i^{\,2}}$ in Eq.~(\ref{CG.me.formula}) has no ${\bm e}_i$-dependence, we can omit it, simplifying Eq.~(\ref{CG.me.formula}) to 
\begin{align}
\langle \Psi^{L'M'\pi'}_{l'}(u',A') |{\cal O}_{\kappa \mu}| \Psi_{l}^{LM\pi}(u,A)\rangle
& \equiv \frac{1}{\sqrt{2L'+1}}\langle LM\, \kappa \mu|L'M'\rangle \langle 
\Psi^{L'\pi'}_{l'}(u',A')||{\cal O}_{\kappa}||\Psi_{l}^{L\pi}(u,A)\rangle \nonumber \\
&= {\cal G} \int \cdots \int d{\bm e}_1 \cdots d{\bm e}_4 \, {\cal D}_{l_1\cdots l_4}^{LML'M'}(\alpha_1 \cdots \alpha_4) \, e^{\frac{1}{2}\widetilde{{\bm \upsilon}}Z^{-1}{\bm \upsilon}}F_i(\bm \upsilon),
\label{CG.me.final}
\end{align}
where $\langle 
\Psi^{L'\pi'}_{l'}(u',A')||{\cal O}_{\kappa}||\Psi_{l}^{L\pi}(u,A)\rangle$ is the reduced matrix element of ${\cal O}_{\kappa \mu}$ and  
\begin{align}
{\cal G}&\equiv{\cal F}\Big(\frac{(2\pi)^6}{{\rm det}Z}\Big)^{\frac{3}{2}}=\big(\prod_{i=1}^4 B_{l_i}\big)(\pi {\rm det}\Gamma)^{\frac{9}{2}}({\rm det}\, {\cal U})^{-\frac{3}{2}},
\label{def.cal.G}
\\
{\cal U}&=\left(\begin{array}{cc}
\Gamma-A & 0  \\
0 & \Gamma-A' \\
\end{array}
\right)Z=\left(\begin{array}{cccc}
\Gamma_1-A_1 & 0 & 0 & 0 \\
0 & \Gamma_2-A_2 & 0 & 0 \\
0 & 0 & \Gamma_3-A_3 & 0 \\
0 & 0 & 0 & \Gamma_4-A_4 \\
\end{array}
\right)Z.
\label{def.cal.U}
\end{align}
${\cal F}\Big(\frac{(2\pi)^6}{{\rm det}Z}\Big)^{\frac{3}{2}}$ contains the product 
of three determinants, ${\rm det}(\Gamma-A),\ {\rm det}(\Gamma-A')$, and ${\rm det}Z$. It could happen that some of them are negative depending on the parameters $A$ and/or $A'$. To avoid that problem, one should combine them to construct the matrix ${\cal U}$ and 
calculate its determinant to obtain ${\cal G}$.  
 I  explain in Appendix~\ref{negative.det} using a very simple example where the problem arises and how it is treated. 

In what follows I show how to carry out the integro-differential operation~(\ref{CG.me.final}) for three cases of $F_i(\bm \upsilon)\, (i=1,2,3)$. \\

\noindent
(i) $F_1(\bm \upsilon)$\\
\indent
Let $\frac{1}{2}\widetilde{{\bm \upsilon}}Z^{-1}{\bm \upsilon}$ be expressed by 
\begin{align}
\sum_{1\le i< j\le 4}\rho_{ij}\alpha_i \alpha_j {\bm e}_i\cdot{\bm e}_j, \ \ \ \ \ \ \ 
\rho_{ij}=\widetilde{u_i}(\Gamma_i-A_i)^{-1}\Gamma_iZ^{-1}_{\,ij}\Gamma_j(\Gamma_j-A_j)^{-1}u_j.
\end{align}
Using  the formula (V) of Appendix~\ref{formulas}, I obtain 
\begin{align}
\prod_{i=1}^4{\cal D}^{l_i}(\alpha_i)\, e^{\frac{1}{2}\widetilde{{\bm \upsilon}}Z^{-1}{\bm \upsilon}}F_1(\bm \upsilon)
& \Rightarrow 
\sum_L G(\{\rho_{ij}\};l_1l_2l_3l_4;L)\Big[[Y_{l_1}({\bm e}_1)\!\times\! Y_{l_2}({\bm e}_2)]_L\ \times\ [Y_{l_3}({\bm e}_3)\!\times\! Y_{l_4}({\bm e}_4)]_L\Big]_{00}.
\end{align}
Substituting the result into Eq.~(\ref{CG.me.final}) and using the formula (VII) shows that the matrix element is 
\begin{align}
\delta_{L,L'}\delta_{M,M'}\delta_{\pi, \pi'}\, {\cal G}\frac{(-1)^{l_1+l_2}}{\sqrt{2L+1}}
G(\{\rho_{ij}\};l_1l_2l_3l_4;L).
\label{F1.mat.ele}
\end{align} \\

\noindent
(ii) $F_2(\bm \upsilon)$\\
\indent
Let $\widetilde{{\bm \upsilon}}X{\bm \upsilon}$ be expressed by 
$2\sum_{1\le i< j\le 4} \tau_{ij}\alpha_i \alpha_j {\bm e}_i\cdot{\bm e}_j$ with 
$\tau_{ij}=\widetilde{u_i}(\Gamma_i-A_i)^{-1}\Gamma_i X_{ij}\Gamma_j(\Gamma_j-A_j)^{-1}u_j$. The matrix element  is evaluated from 
\begin{align}
\prod_{i=1}^4{\cal D}^{l_i}(\alpha_i)\, e^{\frac{1}{2}\widetilde{{\bm \upsilon}}Z^{-1}{\bm \upsilon}}F_2(\bm \upsilon)={\cal D}^1(x) \prod_{i=1}^4{\cal D}^{l_i}(\alpha_i)\, e^{\sum_{1\le i< j\le 4}(\rho_{ij}+2x\tau_{ij})\alpha_i \alpha_j {\bm e}_i\cdot{\bm e}_j}.
\end{align}
The desired matrix element is obtained from Eq.~(\ref{F1.mat.ele}) as follows;
\begin{align}
\delta_{L,L'}\delta_{M,M'}\delta_{\pi, \pi'}\, {\cal G}\frac{(-1)^{l_1+l_2}}{\sqrt{2L+1}}{\cal D}^1(x)
G(\{\rho_{ij}+2x\tau_{ij}\};l_1l_2l_3l_4;L).
\end{align}
Here, I obtain ${\cal D}^1(x)G(\{\rho_{ij}+2x\tau_{ij}\};l_1l_2l_3l_4;L)$ using Eqs.~(\ref{D.on.G}) and (\ref{diff.Ps}). \\

\noindent
(iii) $F_3(\bm \upsilon)$\\
\indent
Let $\widetilde{\omega}{\bm \upsilon}$ be expressed by $\sum_{i=1}^4 \gamma_i\alpha_i{\bm e}_i$.  Using Leibnitz rule I start from the following expression;
\begin{align}
\prod_{i=1}^4{\cal D}^{l_i}(\alpha_i)\,e^{\frac{1}{2}\widetilde{{\bm \upsilon}}Z^{-1}{\bm \upsilon}}F_3(\bm \upsilon)
&=\prod_{i=1}^4 \left\{\sum_{\lambda_i=0}^{l_i} \Big({\cal D}^{l_i-\lambda_i}(\alpha_i)
\,e^{\frac{1}{2}\widetilde{{\bm \upsilon}}Z^{-1}{\bm \upsilon}}\Big)\Big({\cal D}^{\lambda_i}(\alpha_i)F_3(\bm \upsilon)\Big)\right\}\nonumber \\
&= \sum_{\lambda_1=0}^{l_1} \sum_{\lambda_2=0}^{l_2} \sum_{\lambda_3=0}^{l_3} \sum_{\lambda_4=0}^{l_4} 
\Big( \prod_{i=1}^4  {\cal D}^{l_i-\lambda_i}(\alpha_i)\,e^{\frac{1}{2}\widetilde{{\bm \upsilon}}Z^{-1}{\bm \upsilon}} \Big) \Big( \prod_{i=1}^4  {\cal D}^{\lambda_i}(\alpha_i) F_3(\bm \upsilon)\Big).
\end{align}
The formulas (V) and (VI) make it possible to obtain 
\begin{align}
\prod_{i=1}^4{\cal D}^{l_i}(\alpha_i)\,e^{\frac{1}{2}\widetilde{{\bm \upsilon}}Z^{-1}{\bm \upsilon}}F_3(\bm \upsilon)
&\Rightarrow \sum_{\lambda_1=0}^{l_1} \sum_{\lambda_2=0}^{l_2} \sum_{\lambda_3=0}^{l_3} \sum_{\lambda_4=0}^{l_4} \sum_{\bar L} G(\{ \rho_{ij} \}; l_1\!-\!\lambda_1\ l_2\!-\!\lambda_2\ l_3\!-\!\lambda_3\ l_4\!-\!\lambda_4; \bar{L})\nonumber \\
&\quad \times (4\pi)^{\frac{3}{2}}\sqrt{(2\kappa+1)!}\, \delta(\textstyle{\sum_{i=1}^4} \lambda_i,\kappa)
\Big(\prod_{i=1}^4 \frac{\gamma_i^{\lambda_i}}{\sqrt{(2\lambda_i+1)!}}\Big)\nonumber \\
&\quad \times \Big[[Y_{l_1-\lambda_1}({\bm e}_1)\!\times\! Y_{l_2-\lambda_2}({\bm e}_2)]_{\bar L}\ \times\ [Y_{l_3-\lambda_3}({\bm e}_3)\!\times\! Y_{l_4-\lambda_4}({\bm e}_4)]_{\bar L}\Big]_{00}
\nonumber \\
&\quad \times 
[Y_{\lambda_1}({\bm e}_1)\times Y_{\lambda_2}({\bm e}_2)\times Y_{\lambda_3}({\bm e}_3)
\times Y_{\lambda_4}({\bm e}_4)]_{\kappa \mu}.
\end{align}
With the help of the formula (VII) I obtain the desired matrix element as
\begin{align}
&{\cal G} (-1)^{l_1+l_2} \sqrt{\frac{2\kappa+1}{2L'+1}}\langle LM \kappa \mu|L'M' \rangle
(4\pi)^{\frac{3}{2}}\sqrt{(2\kappa+1)!}\nonumber \\
& \quad \times \sum_{\lambda_1=0}^{l_1} \sum_{\lambda_2=0}^{l_2} \sum_{\lambda_3=0}^{l_3} \sum_{\lambda_4=0}^{l_4} \delta(\textstyle{\sum_{i=1}^4} \lambda_i,\kappa) 
\Big(\prod_{i=1}^4 \frac{\gamma_i^{\lambda_i}}{\sqrt{(2\lambda_i+1)!}}\Big)\nonumber \\
& \quad \times \sum_{{\bar L}={\bar L}_{\rm min}}^{{\bar L}_{\rm max}} G(\{ \rho_{ij} \}; l_1\!-\!\lambda_1\ l_2\!-\!\lambda_2\ l_3\!-\!\lambda_3\ l_4\!-\!\lambda_4; \bar{L})\, \Lambda(\lambda_1 \lambda_2 \lambda_3 \lambda_4,l_1l_2l_3l_4, \bar{L};\kappa  L L'),
\end{align}
where 
\begin{align}
&{\bar L}_{\rm min}={\rm max}\,(|L-(\lambda_1+\lambda_2)|,\, |L'-(\lambda_3+\lambda_4)|,\, |l_1-\lambda_1-(l_2-\lambda_2)|,\, |l_3-\lambda_3-(l_4-\lambda_4)|),\nonumber \\ 
&{\bar L}_{\rm max}={\rm min}\,(L+\lambda_1+\lambda_2,\, L'+\lambda_3+\lambda_4,\, l_1-\lambda_1+l_2-\lambda_2,\, l_3-\lambda_3+l_4-\lambda_4). 
\end{align}

The cases of $F_4(\bm \upsilon)$ and $F_5(\bm \upsilon)$ can be treated following the case (ii), and the matrix elements are obtained using Eqs.~(\ref{diff.Ps}) and (\ref{DxDyonP}), respectively.

\section{Summary and discussions}
\label{summary}

Several low-lying levels of $^{16}$O play an important role in the $^{12}$C$(\alpha,\gamma)^{16}$O radiative-capture reaction.
I have developed a fully microscopic $^{12}$C$+\alpha$ model to describe those states. Here, the wave function of $^{12}$C described by $3\,\alpha$-particle model can be excited to, {\it e.g.} the $2^+$ state and it has freedom to be distorted from the one in an isolated system. Obviously the present model can describe 
the $^{12}$C$-\alpha$ asymptotics  very well because it is sensitive to the parameter $A_2$ of Eq.~(\ref{basis.func}). 
 Both of the $E1$ and $E2$ transitions play a key role in the radiative-capture reaction. Especially the description of the $E1$ transition is crucially important because the leading-order term of the isovector $E1$ operator has no contribution if the isospin-impurity 
component is missing. This calls for due consideration on both the isospin-impurity components of the participating nuclei as well as the isoscalar $E1$ operators of the next order. I proposed that the isospin-forbidden $E1$ 
transition could be accounted for through the $T=1$ impurity component of $\alpha$-particle. It should be noted that its component is not an adjustable parameter but in principle determined a priori. 

The wave functions of the low-lying levels of $^{16}$O are assumed to be 
expressed in terms of a 
combination of the correlated-Gaussian basis functions.  I have provided all the formulas that make it possible to 
obtain the needed matrix elements analytically. Because each basis function 
contains several variational parameters, a stochastic  search for the basis functions appears to be inevitable to obtain a converged solution as well as to reduce a computational load~\cite{varga95,book98,mitroy13}. 

It may be hard, however, to reproduce 
consistently both the energies and some important physical observables of all the relevant states.  This may be partly because all of them are assumed to be described by $\alpha$-cluster wave functions and therefore the nucleon-nucleon interaction to be employed contains only central forces. To be consistent with the model, the interaction should reproduce the low-energy 
$\alpha$-$\alpha$ scattering data. For example, Volkov potential~{\cite{volkov65}} is most often used as such an effective potential. In order to  determine the state $\Psi_{LM\pi}$ under the condition, it appears to be reasonable 
to attempt at minimizing the expectation value of $S(\lambda)$,
\begin{align}
S(\lambda)=H+\lambda R^2,
\end{align}
rather than minimizing the  expectation value of the Hamiltonian $H$~\cite{suzuki24}.
Here, $\lambda$ is a parameter and $R^2$ is the mean square radius. Given $\lambda$, I search for such a solution
that minimizes the expectation value of $S(\lambda)$. Let the corresponding solution be $\Psi_{LM\pi}({\lambda})$. At the same time I calculate the energy, $\langle H\rangle_{\lambda}=\langle \Psi_{LM\pi}({\lambda})|H|\Psi_{LM\pi}({\lambda})\rangle$, as well as an important physical observable, $\langle {\cal O}\rangle_{\lambda}$, using that solution. An appropriate value of $\lambda$ is selected by examining both  
$\langle H\rangle_{\lambda}$ and $\langle {\cal O}\rangle_{\lambda}$ as a function of $\lambda$. For instance, in the case of 
determining the $0_1^+$ state, $\langle {\cal O}\rangle_{\lambda}$ could be the mean square radius of the ground state of $^{16}$O that is experimentally known. 
Once the ground state wave function of $^{16}$O is set, the $1_1^-$ state could be determined as follows. By choosing 
$\langle {\cal O}\rangle_{\lambda}$ to be the $E1$ transition probability to the ground state, 
I examine both $\langle H\rangle_{\lambda}$ and $\langle {\cal O}\rangle_{\lambda}$ as 
a function of $\lambda$ and select the best value of $\lambda$. The 
corresponding $\Psi_{1M-}({\lambda})$ is chosen to be the $1_1^-$ wave function. 

A great merit of the above approach compared to the usual minimization of the expectation value of $H$ is that one can apply it to a broad resonance state such as the $1_2^-$ state at 9.59 MeV excitation energy. It is broad because of the $\alpha$-decay but its $E1$ decay width to the ground state is about 2.5$\times 10^{-2}$ eV~\cite{tilley93,database}. With a positive $\lambda$ value $\Psi_{1M-}({\lambda})$ does not diverge at large distances but can be confined, so that there may be a chance that $\Psi_{1M-}({\lambda})$ predicts the $E1$ decay width close to the observed value. 

For the case of $L^{\pi}=0^+,\, 2^+$, and $1^-$, I have to generate two solutions $\Psi_{LM\pi}({\lambda_1})$ and $\Psi_{LM\pi}({\lambda_2})$.  Since $\lambda_1$ and $\lambda_2$ are different in general, they are not orthogonal each other. This problem has to be fixed suitably in order to determine the two states 
with the same $L\pi$. 

Once the five subthreshold states and two states above $^{12}$C$+\alpha$ threshold of $^{16}$O are set up, one will be ready for a microscopic calculation of the $^{12}$C$(\alpha,\gamma)^{16}$O radiative-capture cross section using, {\it e.g} a microscopic $R$-matrix theory~\cite{Rmatrix10}. It may be possible that one  can predict not only the total cross section but also each of the partial cross sections separately, depending on the different path of the various capture processes.   

A numerical achievement with the proposed microscopic theory appears to be very useful because the theory especially can 
take account of the isospin-forbidden $E1$ transition in a simple way and also because no calculations of that kind have ever been done 
within the microscopic $4\,\alpha$ cluster-model. It should be recalled, however, that realistic nucleon-nucleon ineteractions contain strong tensor components as well as short-ranged repulsion. Though the $(0s)^4$ wave function assumed for $\alpha$ particle is a reasonable approximation, the $(0s)^4$ component 
predicted with such realistic interactions is at most 80$\,\%$, and in addition the $D$-state probability of $\alpha$ particle amounts to more than 10$\,\%$~\cite{horiuchi14}. On the other hand, the value of $\epsilon$ estimated in Ref.~\cite{suzuki21} is consistent with that given by realistic potentials~\cite{wiringa}. The theory formulated here will serve as a reasonable approximation provided that the low-lying levels of $^{16}$O can be well described by a microscopic $4\,\alpha$ cluster-model.

\appendix

\section{Negative determinants}
\label{negative.det}

In order to understand how the negative determinant problem arises and how it is 
avoided, let 
us take a simple example, the evaluation of the overlap 
of CG's in two ways. One is a direct integration,
\begin{align}
O(A',A)=\langle e^{-\frac{1}{2}\tilde{\bm x}A'\bm x}|e^{-\frac{1}{2}\tilde{\bm x}A\bm x}\rangle=
\Big(\frac{(2\pi)^3}{{\rm det}(A+A')}\Big)^{\frac{3}{2}}.
\label{direct}
\end{align}
Here, ${\bm x}$ stands for the set of 3-dimensional vectors, $\{{\bm x}_1, {\bm x}_2, {\bm x}_3\}$, and $A$ and $A'$ are $3\times 3$ real, symmetric positive-definite matrices.  Sine $A+A'$ is positive-definite, one obtains $O(A',A)$ without any problem. 
The other is an indirect way by expressing the 
CG  via the GWP in the form ({\it cf}. Eqs.~(\ref{xtos}) and (\ref{def.Qv}))
\begin{align}
e^{-\frac{1}{2}\tilde{\bm x}A\bm x}&=C_A\int d{\bm s}\, e^{-\frac{1}{2} \tilde{\bm s} Q_A\bm s}\psi(\bm s, \bm x),
\label{stox}
\\
\psi(\bm s, \bm x)&=\pi^{-\frac{9}{4}}({\rm det}\, \Gamma)^{\frac{3}{4}} \, e^{-\frac{1}{2}\widetilde{\bm x}\Gamma \bm x+\widetilde{\bm x}\Gamma \bm s-\frac{1}{2}\widetilde{\bm s}\Gamma \bm s},
\end{align}
with  
\begin{align}
&C_A= \Big(\frac{{\rm det \Gamma}}{4\pi}\Big)^{\frac{9}{4}}[{\rm det}\, (\Gamma-A)]^{-\frac{3}{2}},\ \ \ \ \ Q_A=\Gamma(\Gamma-A)^{-1}\Gamma - \Gamma.
\label{G-A}
\end{align}
Here, ${\bm s}$ denotes the set of 3-dimensional variables, $\{{\bm s}_1, {\bm s}_2, {\bm s}_3\}$, and $\Gamma=(\gamma_i \delta_{i,j})$ is a $3\times 3$ diagonal matrix with $\gamma_i>0$. The integration of Eq.~(\ref{stox}) converges only when $Q_A+\Gamma$ is positive-definite. That is, 
$\Gamma-A$ has to be positive-definite. Let us ignore this condition temporarily. 

By using 
\begin{align}
\langle \psi(\bm s',\bm x)|\psi({\bm s},\bm x)\rangle=
e^{-\frac{1}{4}\widetilde{(\bm s-{\bm s}')}\Gamma (\bm s-{\bm s}')},
\end{align}
and introducing ${\bm t}=
\left(\begin{array}{c}
{\bm s}  \\
{\bm s}' \\
\end{array}
\right)$, I find that the overlap takes the form
\begin{align}
O(A',A)&= C_AC_{A'}\iint d{\bm s} d{\bm s'} 
e^{-\frac{1}{2}\tilde{\bm s}Q_A\bm s-\frac{1}{2}\tilde{\bm s'}Q_{A'}{\bm s}'}
\langle \psi(\bm s',\bm x)|\psi({\bm s},\bm x)\rangle \nonumber \\
&=C_{A}C_{A'}\int d{\bm t}\,  e^{-\frac{1}{2}\tilde{\bm t}Z \bm t} =C_{A} C_{A'} 
\Big(\frac{(2\pi)^6}{{\rm det}Z}\Big)^{\frac{3}{2}},
\label{indirect}
\end{align}
where $Z$ is a $6\times 6$ symmetric matrix defined by 
\begin{align}
Z&=\Big(\begin{array}{cc}
Q_A+\frac{1}{2}\Gamma & -\frac{1}{2}\Gamma \\
-\frac{1}{2}\Gamma & Q_{A'}+\frac{1}{2}\Gamma \\  
\end{array}
\Big)
=\Big(
\Big(\begin{array}{cc}
M^{-1} & 0 \\
0 & M'^{-1} \\  
\end{array}
\Big)
-\frac{1}{2}
\Big(\begin{array}{cc}
1 & 1 \\
1 & 1 \\  
\end{array}
\Big)
\Big)
\Big(\begin{array}{cc}
\Gamma & 0 \\
0 & \Gamma \\  
\end{array}
\Big).
\end{align}
Here, $M$ and $M'$ are defined by
\begin{align}
M=(\Gamma -A)\Gamma^{-1},\ \ \ \ \ 
M'=(\Gamma -A')\Gamma^{-1}.
\label{def.MM'}
\end{align}
Note that the complex conjugation of $C_{A'}$ is not taken in Eq.~(\ref{indirect}) because $\psi(\bm s, \bm x)$ is assumed to be real. 

A question is: {\it  The sign of ${\rm det}(\Gamma -A)$ and/or that of ${\rm det}(\Gamma -A')$  may be negative. Is Eq.~(\ref{indirect}) equal to Eq.~(\ref{direct})?} 
An answer to this question crucially depends on whether the following equation holds the case or not:
\begin{align}
{\rm det}\Big(\frac{A+A'}{2}\Big)&={\rm det}(\Gamma -A) \, {\rm det}(\Gamma -A')  \, ({\rm det}\Gamma)^{-3} \, {\rm det}Z\nonumber \\
&= {\rm det}\, \Gamma \, {\rm det}M \, {\rm det}M'\, 
{\rm det}\Big(
\Big(\begin{array}{cc}
M^{-1} & 0 \\
0 & M'^{-1} \\  
\end{array}
\Big)
-\frac{1}{2}
\Big(\begin{array}{cc}
1 & 1 \\
1 & 1 \\  
\end{array}
\Big)
\Big).
\label{eq.tobeproved}
\end{align}
It is better to rewrite this equation as follows. By replacing ${\rm det}M \, {\rm det}M'$ with ${\rm det} \Big(\begin{array}{cc}
M & 0 \\
0 & M' \\  
\end{array}\Big)$, Eq.~(\ref{eq.tobeproved}) reduces to 
\begin{align}
{\rm det}\Big(\frac{A+A'}{2}\Big)
&={\rm det \Gamma} \, {\rm det}
\Big(
1-\frac{1}{2}
\Big(\begin{array}{cc}
M & M \\
M' & M' \\  
\end{array}
\Big)
\Big).
\label{eq.test}
\end{align} 
Using $\frac{1}{2}(A+A')=(1-\frac{1}{2}(M+M'))\Gamma$ from Eq.~(\ref{def.MM'}) and 
substituting it to the left-hand side of the above equation, 
Eq.~(\ref{eq.test}) reduces to 
\begin{align}
{\rm det}\Big(1-\frac{1}{2}(M+M')\Big) = {\rm det}
\Big(
1-\frac{1}{2}
\Big(\begin{array}{cc}
M & M \\
M' & M' \\  
\end{array}
\Big)
\Big).
\label{dettriple}
\end{align}
It is possible to prove that Eq.~(\ref{dettriple}) is in fact valid as follows:
\begin{align}
{\rm det}
\Big(
1-\frac{1}{2}
\Big(\begin{array}{cc}
M & M \\
M' & M' \\  
\end{array}
\Big)
\Big)
&={\rm det}\Big(\begin{array}{cc}
1-\frac{1}{2}M & -\frac{1}{2}M \\
-\frac{1}{2}M' & 1-\frac{1}{2}M'\\
\end{array}
\Big)
={\rm det}\Big(\begin{array}{cc}
1 & -\frac{1}{2}M \\
-1 & 1-\frac{1}{2}M' \\
\end{array}
\Big)\nonumber \\
&={\rm det}\Big(\begin{array}{cc}
1 & -\frac{1}{2}M \\
0 & 1-\frac{1}{2}M-\frac{1}{2}M' \\
\end{array}
\Big)
={\rm det}
\Big(1-\frac{1}{2}(M+M')\Big).
\end{align}

The above analysis suggests that I 
should not multiply three terms, $({\rm det}M)^{\frac{3}{2}}$, $({\rm det} M')^{\frac{3}{2}}$, and 
$({\rm det}Z)^{\frac{3}{2}}$, but combine them to the single term,  $\Big[{\rm det}\, 
\Big(\big(\begin{array}{cc}
M & 0 \\
0 & M' \\  
\end{array}\big) Z\Big)\Big]^{\frac{3}{2}}$. The negative determinant problem 
arises because I calculate the CG matrix element through the GWP used in 
the microscopic cluster-model. The problem is fortunately dodged by introducing ${\cal G}$ of 
Eq.~(\ref{def.cal.G}) and calculating it via the matrix ${\cal U}$ of Eq.~(\ref{def.cal.U}).

\section {Formulas for angular-momentum projection}
\label{formulas}

The matrix elements for $F_i({\bm \upsilon})\, (i=1,2,3)$ in Sec.~\ref{calculation.m.e.} indicate that one needs formulas to couple several 
spherical harmonics and to integrate the product of several spherical harmonics. Some of the formulas were previously derived in Ref.~\cite{suzuki08}. In this 
appendix I prepare the formulas needed to derive the matrix elements for 
all of  $F_i({\bm \upsilon})$'s. 
A key ingredient to derive the formulas is the stretched coupling for 
the spherical harmonics of the same arguments, as shown in Eqs.~(\ref{stretched.coupling}) and (\ref{YY.power}).  
In what follows I use the angular momentum recoupling coefficients, $6j$ symbol or Racah coefficient and $9j$ symbol in unitary form. See, {\it e.g.} Ref.~\cite{book98} for their definitions and properties. 

\begin{flalign}
&{\rm (I)} \quad [Y_{a}({\bm e}_1)\!\times \!Y_{a}({\bm e}_3)]_{00}\ 
   [Y_{b}({\bm e}_1)\!\times \! Y_{b}({\bm e}_4)]_{00}\ 
   [Y_{c}({\bm e}_2)\!\times \!Y_{c}({\bm e}_3)]_{00}\ 
   [Y_{d}({\bm e}_2)\!\times \!Y_{d}({\bm e}_4)]_{00}\nonumber &\\ 
&\qquad \Rightarrow 
\sum_{L=L_{\rm min}}^{a+b+c+d} X(abcd;L) \Big[[Y_{a+b}(\bm e_1)\! \times\!  Y_{c+d}(\bm e_2)]_L \, \times\,  
[Y_{a+c}(\bm e_3)\!\times\! Y_{b+d}(\bm e_4)]_L\Big]_{00},&
\label{AppI.1}
\end{flalign}
\quad where $L_{\rm min}={\rm max}(|a+b-(c+d)|,\, |a+c-(b+d)|)$ and 
\begin{align}
X(abcd;L)&=\sqrt{\frac{2L\!+\!1}{(2a\!+\!1)(2b\!+\!1)(2c\!+\!1)(2d\!+\!1)}} 
\, C(a\,b;a\!+\!b)\,C(c\,d;c\!+\!d)\,C(a\,c;a\!+\!c)\,C(b\,d;b\!+\!d)
\left[
\begin{array}{ccc}
a & b & a\!+\!b \\
c & d & c\!+\!d \\
a\!+\!c & b\!+\!d & L\\
\end{array}
\right].
\end{align}

\noindent
{\it Proof.\ \ } The coefficient $X(abcd;L)$ is derived using the $9j$ symbol as shown below: 
\begin{align}
&[Y_{a}({\bm e}_1)\!\times \!Y_{a}({\bm e}_3)]_{00}\ 
   [Y_{b}({\bm e}_1)\!\times \! Y_{b}({\bm e}_4)]_{00}\ 
   [Y_{c}({\bm e}_2)\!\times \!Y_{c}({\bm e}_3)]_{00}\ 
   [Y_{d}({\bm e}_2)\!\times \!Y_{d}({\bm e}_4)]_{00}\nonumber \\
& \Rightarrow \left[
\begin{array}{ccc}
a & a & 0 \\
b & b & 0 \\
a\!+\!b & a\!+\!b & 0\\
\end{array}
\right]C(a\,b;a\!+\!b)\,\Big[Y_{a+b}(\bm e_1)\ \times\  
[Y_{a}(\bm e_3)\! \times \! Y_{b}(\bm e_4)]_{a+b}\Big]_{00}\nonumber \\
&\ \times \left[
\begin{array}{ccc}
c & c & 0 \\
d & d & 0 \\
c\!+\!d & c\!+\!d & 0\\
\end{array}
\right]C(c\,d;c\!+\!d)\,\Big[Y_{c+d}(\bm e_2)\ \times\  
[Y_{c}(\bm e_3)\!\times \!Y_{d}(\bm e_4)]_{c+d}\Big]_{00}\nonumber \\
&=\sum_L \sqrt{\frac{(2(a+b)+1)(2(c+d)+1)}{(2a+1)(2b+1)(2c+1)(2d+1)}} 
C(a\,b;a\!+\!b)\,C(c\,d;c\!+\!d)\nonumber \\
&\ \times 
\left[
\begin{array}{ccc}
a+b & a+b & 0 \\
c+d & c+d & 0 \\
L & L & 0\\
\end{array}
\right]
\Big[[Y_{a+b}(\bm e_1)\! \times \! Y_{c+d}(\bm e_2)]_L \ \times \ 
\big[[Y_{a}(\bm e_3)\!\times \!Y_{b}(\bm e_4)]_{a+b}\times 
[Y_{c}(\bm e_3)\!\times \!Y_{d}(\bm e_4)]_{c+d}\big]_L\Big]_{00}\nonumber \\
&\Rightarrow \sum_L \sqrt{\frac{2L\!+\!1}{(2a\!+\!1)(2b\!+\!1)(2c\!+\!1)(2d\!+\!1)}} 
C(a\,b;a\!+\!b)\,C(c\,d;c\!+\!d)\,C(a\,c;a\!+\!c)\,C(b\,d;b\!+\!d)\nonumber \\
&\ \times 
\left[
\begin{array}{ccc}
a & b & a\!+\!b \\
c & d & c\!+\!d \\
a\!+\!c & b\!+\!d & L\\
\end{array}
\right]
\Big[[Y_{a+b}(\bm e_1)\!\times\! Y_{c+d}(\bm e_2)]_L \ \times\  
[Y_{a+c}(\bm e_3)\!\times\! Y_{b+d}(\bm e_4)]_L\Big]_{00}.
\end{align}

\begin{flalign}
&{\rm (II)}\quad 
\Big[[Y_{a}(\bm e_1)\!\times\! Y_{b}(\bm e_2)]_{\ell} \ \times\ 
[Y_{c}(\bm e_3)\!\times\! Y_{d}(\bm e_4)]_{\ell}\Big]_{00}\ 
\Big[[Y_{\alpha}(\bm e_1)\!\times\! Y_{\beta}(\bm e_2)]_{\lambda} \ \times\  
[Y_{\gamma}(\bm e_3)\!\times\! Y_{\delta}(\bm e_4)]_{\nu}\Big]_{\kappa \mu}\nonumber &\\
&\qquad \ \Rightarrow \sum_{L={L_{\rm min}}}^{L_{\rm max}}\sum_{L'={L'_{\rm min}}}^{L'_{\rm max}} Y(abcd, \alpha \beta \gamma \delta, \ell \lambda \nu \kappa; LL')
\Big[[Y_{a+\alpha}(\bm e_1)\!\times\! Y_{b+\beta}(\bm e_2)]_{L} \ \times\  
[Y_{c+\gamma}(\bm e_3)\!\times\! Y_{d+\delta}(\bm e_4)]_{L'}\Big]_{\kappa \mu},
\label{AppII.1}
\end{flalign}
\qquad  where 
\begin{flalign}
L_{\rm min}&={\rm max}(|l-\lambda|,\, |a+\alpha-(b+\beta)|),\ \ \ \ 
L_{\rm max}={\rm min}(l+\lambda,\,  a+\alpha+b+\beta),\nonumber \\
 L'_{\rm min}&={\rm max}(|l-\nu|,\, |c+\gamma-(d+\delta)|), \ \ \ \ \
L'_{\rm max}={\rm min}(l+\nu,\,  c+\gamma+d+\delta),
\end{flalign}
\qquad  and 
\begin{flalign}
&\qquad \qquad Y(abcd, \alpha \beta \gamma \delta, \ell \lambda \nu \kappa; LL')
=C(a\, \alpha;a\!+\!\alpha)\,C(b\, \beta;b\!+\!\beta)\,C(c\, \gamma;c\!+\!\gamma)\,C(d\, \delta;d\!+\!\delta)
\nonumber &\\
&\qquad \qquad \qquad \qquad \qquad \qquad \qquad \quad \ \  \times 
\left[
\begin{array}{ccc}
\ell & \ell & 0 \\
\lambda & \nu & \kappa \\
L & L' & \kappa\\
\end{array}
\right]
\left[
\begin{array}{ccc}
a & b & \ell \\
\alpha & \beta & \lambda \\
a\!+\!\alpha & b\!+\!\beta & L\\
\end{array}
\right]
\left[
\begin{array}{ccc}
c & d & \ell \\
\gamma & \delta & \nu \\
c\!+\!\gamma & d\!+\!\delta & L'\\
\end{array}
\right].&
\label{def.Ycoef}
\end{flalign}

\noindent
{\it Proof.\ \ }
\begin{align}
&\Big[[Y_{a}(\bm e_1)\!\times\! Y_{b}(\bm e_2)]_{\ell} \ \times\ 
[Y_{c}(\bm e_3)\!\times\! Y_{d}(\bm e_4)]_{\ell}\Big]_{00}\ 
\Big[[Y_{\alpha}(\bm e_1)\!\times\! Y_{\beta}(\bm e_2)]_{\lambda} \ \times\  
[Y_{\gamma}(\bm e_3)\!\times\! Y_{\delta}(\bm e_4)]_{\nu}\Big]_{\kappa \mu}\nonumber \\
&=\sum_{LL'}\left[
\begin{array}{ccc}
\ell & \ell & 0 \\
\lambda & \nu & \kappa \\
L & L' & \kappa\\
\end{array}
\right]
\Big[\big[[Y_{a}(\bm e_1)\!\times\! Y_{b}(\bm e_2)]_{\ell} \times [Y_{\alpha}(\bm e_1)\!\times\! Y_{\beta}(\bm e_2)]_{\lambda}\big]_{L} \times 
\big[[Y_{c}(\bm e_3)\!\times\! Y_{d}(\bm e_4)]_{\ell} \times [Y_{\gamma}(\bm e_3)\!\times\! Y_{\delta}(\bm e_4)]_{\nu}\big]_{L'}\Big]_{\kappa \mu}.
\label{II.1}
\end{align}
The coupling of $Y's$ of the same arguments simplifies to 
\begin{align}
&\big[[Y_{a}(\bm e_1)\!\times\! Y_{b}(\bm e_2)]_{\ell} \times [Y_{\alpha}(\bm e_1)\!\times\! Y_{\beta}(\bm e_2)]_{\lambda}\big]_{L} \Rightarrow 
\left[
\begin{array}{ccc}
a & b & \ell \\
\alpha & \beta & \lambda \\
a\!+\!\alpha & b\!+\!\beta & L\\
\end{array}
\right]C(a\,\alpha;a+\alpha)\,C(b\,\beta;b+\beta)[Y_{a+\alpha}(\bm e_1)\!\times\! Y_{b+\beta}(\bm e_2)]_{L},\nonumber \\
&\big[[Y_{c}(\bm e_3)\!\times\! Y_{d}(\bm e_4)]_{\ell} \times [Y_{\gamma}(\bm e_3)\!\times\! Y_{\delta}(\bm e_4)]_{\nu}\big]_{L'} \Rightarrow
\left[
\begin{array}{ccc}
c & d & \ell \\
\gamma & \delta & \nu \\
c\!+\!\gamma & d\!+\!\delta & L'\\
\end{array}
\right]C(c\,\gamma;c+\gamma)\,C(d\,\delta;d+\delta)[Y_{c+\gamma}(\bm e_3)\!\times\! Y_{d+\delta}(\bm e_4)]_{L'}.
\label{II.2}
\end{align}
Substitution of Eq.~(\ref{II.2}) into Eq.~(\ref{II.1}) completes the proof.
Note that the 9$j$ symbol of the following type reduces to the Racah coefficient $U$: 
\begin{align}
 \left[
\begin{array}{ccc}
j_1 & j_1 & 0 \\
j_3 & j_4 & J \\
J_{13} & J_{24} & J\\
\end{array}
\right] =\sqrt{ \frac{2J_{13}+1}{(2j_1+1)(2j_3+1)}} U(J_{13}j_1Jj_4;j_3J_{24}).
\label{9jto6j}
\end{align}

\begin{flalign}
&{\rm (III)}\quad  
  [Y_{a}(\bm e_1)\!\times\! Y_{a}(\bm e_2)]_{00}\ 
   [Y_{b}({\bm e}_1)\!\times \!Y_{b}({\bm e}_3)]_{00}\
   [Y_{c}({\bm e}_1)\!\times \! Y_{c}({\bm e}_4)]_{00} \nonumber &\\ 
&\quad  \quad  \times  [Y_{d}({\bm e}_2)\!\times \!Y_{d}({\bm e}_3)]_{00}\ 
   [Y_{e}({\bm e}_2)\!\times \!Y_{e}({\bm e}_4)]_{00}\ 
   [Y_{f}(\bm e_3)\!\times\! Y_{f}(\bm e_4)]_{00}\nonumber &\\
&\quad \quad \ \ \ \Rightarrow \sum_{L=L_{\rm min}}^{L_{\rm max}} Z(abcdef; L) \Big[[Y_{a+b+c}({\bm e}_1)\!\times\! Y_{a+d+e}({\bm e}_2)]_L\ \times\ [Y_{f+b+d}({\bm e}_3)\!\times\! Y_{f+c+e}({\bm e}_4)]_L\Big]_{00},
\end{flalign}
\qquad where $L_{\rm min}={\rm max}(|b+c-(d+e)|,\, |b+d-(c+e)|),\ L_{\rm max}=b+c+d+e$, and 
\begin{flalign}
&\qquad Z(abcdef;L)=X(bcde;L)\sqrt{\frac{(2(a\!+\!b\!+\!c)+1)(2(f\!+\!b\!+\!d)+1)}{(2a\!+\!1)(2(b\!+\!c)+1)(2f\!+\!1)(2(b+d)\!+\!1)}}\nonumber &\\
&\qquad \qquad \qquad \qquad \times C(a\  b\!+\!c;a\!+\!b\!+\!c)\,C(a\  d\!+\!e;a\!+\!d\!+\!e)\, C(f\  b\!+\!d;f\!+\!b\!+\!d)\, C(f\ \ c\!+\!e;f\!+\!c\!+\!e)\nonumber &\\
&\qquad \qquad \qquad \qquad \times U(a\!+\!b\!+\!c\ \, a\ \, L\ \, d\!+\!e;b\!+\!c\ \, a\!+\!d\!+\!e)\,
        U(f\!+\!b\!+\!d\ \, f\ \, L\ \, c\!+\!e;b\!+\!d\ \, f\!+\!c\!+\!e).
\label{def.Z}
\end{flalign}

\noindent
{\it Proof.\ \ } From (I) I obtain 
\begin{align}
&[Y_{b}({\bm e}_1)\!\times \!Y_{b}({\bm e}_3)]_{00}\
   [Y_{c}({\bm e}_1)\!\times \! Y_{c}({\bm e}_4)]_{00}\ 
   [Y_{d}({\bm e}_2)\!\times \!Y_{d}({\bm e}_3)]_{00}\ 
   [Y_{e}({\bm e}_2)\!\times \!Y_{e}({\bm e}_4)]_{00} \nonumber \\
&\Rightarrow 
\sum_L X(bcde;L) \Big[[Y_{b+c}(\bm e_1)\! \times\!  Y_{d+e}(\bm e_2)]_L \, \times\,  [Y_{b+d}(\bm e_3)\!\times\! Y_{c+e}(\bm e_4)]_L\Big]_{00}.
\end{align}
Coupling the remaining terms, $[Y_{a}(\bm e_1)\!\times\! Y_{a}(\bm e_2)]_{00}\ 
[Y_{f}(\bm e_3)\!\times\! Y_{f}(\bm e_4)]_{00}$, with the above term 
and using (II), I obtain   
\begin{align}
&[Y_{a}(\bm e_1)\!\times\! Y_{a}(\bm e_2)]_{00}\ 
[Y_{f}(\bm e_3)\!\times\! Y_{f}(\bm e_4)]_{00}\ 
\Big[[Y_{b+c}(\bm e_1)\! \times\!  Y_{d+e}(\bm e_2)]_L \, \times\,  
[Y_{b+d}(\bm e_3)\!\times\! Y_{c+e}(\bm e_4)]_L\Big]_{00}\nonumber \\
&\Rightarrow Y(aaff,b\!+\!c\ d\!+\!e\ b\!+\!d\ c\!+\!e,0LL0;LL)
\Big[[Y_{a+b+c}({\bm e}_1)\!\times \!Y_{a+d+e}({\bm e}_2)]_L \times [Y_{f+b+d}({\bm e}_3)\! \times \!Y_{f+c+e}({\bm e}_4)]_L\Big]_{00}.
\end{align}
Here, the $Y$ coefficient is obtained from 
Eq.~(\ref{def.Ycoef}), in which the 9$j$ symbols are replaced by the Racah coefficients using Eq.~(\ref{9jto6j}), confirming (III) together with Eq.~(\ref{def.Z}).
\\
\begin{flalign}
&{\rm (IV)}\quad 
\int \cdots \int d{\bm e}_1 \cdots d{\bm e}_4  [Y_{l_3}({\bm e}_3)\!\times \! Y_{l_4}({\bm e}_4)]^*_{L'M'} [Y_{l_1}({\bm e}_1)\!\times \! Y_{l_2}({\bm e}_2)]_{LM} \prod_{1\le i< j\le 4}[Y_{n_{ij}}({\bm e}_i)_{00}\!\times\! Y_{n_{ij}}({\bm e}_j)]_{00}\nonumber &\\
& \qquad \quad \Rightarrow \frac{(-1)^{l_1+l_2}}{\sqrt{2L+1}}\delta_{LL'}\delta_{MM'} Z(n_{12}n_{13}n_{14}n_{23}n_{24}n_{34}; L)\nonumber &\\
& \qquad \quad \ \ \times \delta(n_{12}+n_{13}+n_{14},l_1)\, \delta(n_{12}+n_{23}+n_{24},l_2)\, \delta( n_{13}+n_{23}+n_{34},l_3)\, \delta( n_{14}+n_{24}+n_{34},l_4).&
\end{flalign} 

\noindent
{\it Proof.\ \ }  This is readily proved by using (III) and noting $[Y_{l_3}({\bm e}_3)\!\times \! Y_{l_4}({\bm e}_4)]^*_{L'M'}=(-1)^{l_3+l_4-L'+M'}[Y_{l_3}({\bm e}_3)\!\times \! Y_{l_4}({\bm e}_4)]_{L'\, -M'}$. In what follows, $Z(n_{12}n_{13}n_{14}n_{23}n_{24}n_{34}; L)$ is abbreviated to $Z(\{n_{ij}\}; L)$.
\\
\begin{flalign}
&{\rm (V)}\quad \prod_{k=1}^4{\cal D}^{l_k}(\alpha_k) \, e^{\sum_{1\le i< j\le 4}\rho_{ij}\alpha_i \alpha_j {\bm e}_i \cdot {\bm e}_j}  
\Rightarrow \sum_{L=L_{\rm min}}^{L_{\rm max}} G(\{\rho_{ij}\};l_1l_2l_3l_4;L) \Big[[Y_{l_1}({\bm e}_1)\!\times\! Y_{l_2}({\bm e}_2)]_L\ \times\ [Y_{l_3}({\bm e}_3)\!\times\! Y_{l_4}({\bm e}_4)]_L\Big]_{00},&
\end{flalign}
\qquad  where $L_{\rm min}={\rm max}(|l_1-l_2|,\, |l_3-l_4|),\  L_{\rm max}={\rm min}(l_1+l_2,\,  l_3+l_4)$, and  
\begin{flalign}
& \qquad \qquad G(\{\rho_{ij}\};l_1l_2l_3l_4;L)&\nonumber \\
& \qquad \qquad \qquad \quad =\sum_{n_{12}n_{13}n_{14}n_{23}n_{24}n_{34}} 
Ps(\{n_{ij}\};\{\rho_{ij}\})\, Z(\{n_{ij}\};L)\nonumber &\\
&\qquad \qquad \qquad \quad  \times \delta(n_{12}+n_{13}+n_{14},l_1)\, \delta(n_{12}+n_{23}+n_{24},l_2)\, \delta( n_{13}+n_{23}+n_{34},l_3)\, \delta( n_{14}+n_{24}+n_{34},l_4),&
\label{def.G}
\\
&\qquad  \qquad Ps(\{n_{ij}\};\{\rho_{ij}\})=\prod_{1\le i< j\le 4} \frac{(-1)^{n_{ij}}\sqrt{2n_{ij}+1}}{B_{n_{ij}}}(\rho_{ij})^{n_{ij}}.& 
\label{def.Ps}
\end{flalign}

\noindent
{\it Proof.\ \ } Using Eq.~(\ref{YY.power}) and (III) I obtain 
\begin{align}
 e^{\sum_{1\le i< j\le 4}\rho_{ij}\alpha_i \alpha_j {\bm e}_i \cdot {\bm e}_j}&\Rightarrow
\prod_{1\le i< j\le 4} \left\{\sum_{n_{ij}} \frac{(-1)^{n_{ij}}\sqrt{2n_{ij}+1}}{B_{n_{ij}}}(\rho_{ij}\alpha_i \alpha_j)^{n_{ij}}[Y_{n_{ij}}({\bm e}_i)\!\times \!Y_{n_{ij}}({\bm e}_j)]_{00}\right\}\nonumber \\
&\Rightarrow \sum_L \sum_{n_{12}n_{13}n_{14}n_{23}n_{24}n_{34}} Ps(\{n_{ij}\};\{\rho_{ij}\})\, Z(\{n_{ij}\};L)\nonumber \\
&\quad \times \alpha_1^{n_{12}+n_{13}+n_{14}}\alpha_2^{n_{12}+n_{23}+n_{24}}\alpha_3^{n_{13}+n_{23}+n_{34}}\alpha_4^{n_{14}+n_{24}+n_{34}}\nonumber \\
&\quad \times \Big[[Y_{n_{12}+n_{13}+n_{14}}({\bm e}_1)\!\times\! Y_{n_{12}+n_{23}+n_{24}}({\bm e}_2)]_L\ \times\ [Y_{n_{13}+n_{23}+n_{34}}({\bm e}_3)\!\times\! Y_{n_{14}+n_{24}+n_{34}}({\bm e}_4)]_L\Big]_{00}.
\end{align}
The operation $\prod_{i=1}^4{\cal D}^{l_i}(\alpha_i)$ imposes the following conditions on the non-negative integers $n_{ij}$'s:
\begin{align}
n_{12}+n_{13}+n_{14}=l_1,\ \ \ \  n_{12}+n_{23}+n_{24}=l_2,\ \ \ \  n_{13}+n_{23}+n_{34}=l_3,\ \ \ \  n_{14}+n_{24}+n_{34}=l_4.
\label{psum}
\end{align}
$l_1+l_2-l_3-l_4$ has to be even, otherwise there is no solution for $n_{ij}$'s.  Two among 6 $n_{ij}$'s, {\it e.g.} $n_{12}$ and $n_{13}$, are independent, subject to the constraints
\begin{align}
&n_{12} \geq {\rm max}\,(0,\textstyle{\frac{1}{2}}(l_1+l_2-l_3-l_4)), \ \  n_{13}\geq {\rm max}\,(0,\textstyle{\frac{1}{2}}(l_1-l_2+l_3-l_4)),
  \ \ n_{12}+n_{13} \leq {\rm min}\,(l_1, \textstyle{\frac{1}{2}}(l_1+l_2+l_3-l_4)),
\label{p12p13}
\end{align}
and the rest are determined by 
\begin{align}
&n_{14}=l_1\!-\!n_{12}\!-\!n_{13},\ \ \ \ \ \ \ n_{23}=\textstyle{\frac{1}{2}}(l_1+l_2+l_3-l_4)\!-\!n_{12}\!-\!n_{13}, \nonumber \\
& n_{24}=\textstyle{\frac{1}{2}}(-l_1+l_2-l_3+l_4)+n_{13},\ \ \ \ \ \ \ n_{34}=\textstyle{\frac{1}{2}}(-l_1-l_2+l_3+l_4)\!+\!n_{12}.
\label{p_{ij}}
\end{align} 
Because $l_i$'s are expected to be small integers, only few sets of $\{n_{ij}\}$ satisfy Eq.~(\ref{psum}), contributing to the coefficient $G(\{\rho_{ij}\};l_1l_2l_3l_4;L)$ of Eq.~(\ref{def.G}). \\
It is useful to define ${\cal D}^1(x)G(\{\rho_{ij}+x\tau_{ij}\};l_1l_2l_3l_4;L)$ and ${\cal D}^1(x){\cal D}^1(y)G(\{\rho_{ij}+x\tau_{ij}+y\zeta_{ij}\};l_1l_2l_3l_4;L)$ to perform the operation of Eq.~(\ref{CG.me.final}) for $F_i(\bm \upsilon), \, i=2,4,5$:
\begin{align}
&{\cal D}^1(x)G(\{\rho_{ij}+x\tau_{ij}\};l_1l_2l_3l_4;L)\nonumber \\
&\quad =\sum_{n_{12}n_{13}n_{14}n_{23}n_{24}n_{34}}{\cal D}^1(x) Ps(\{n_{ij}\};\{\rho_{ij}+x\tau_{ij}\})\, Z(\{n_{ij}\};L) \nonumber \\
&\quad \ \ \times \delta(n_{12}+n_{13}+n_{14},l_1)\, \delta(n_{12}+n_{23}+n_{24},l_2)\, \delta( n_{13}+n_{23}+n_{34},l_3)\, \delta( n_{14}+n_{24}+n_{34},l_4).
\label{D.on.G}
\end{align}
where ${\cal D}^1(x) Ps(\{n_{ij}\};\{\rho_{ij}+x\tau_{ij}\})$ is given by 
\begin{align}
&{\cal D}^1(x)Ps(\{n_{ij}\};\{\rho_{ij}+x\tau_{ij}\})\nonumber \\
&\quad =\sum_{1\le k< l\le 4} \frac{(-1)^{n_{kl}}\sqrt{2n_{kl}+1}}{B_{n_{kl}}}
               n_{kl}\tau_{kl}(\rho_{kl})^{n_{kl}-1}
              \prod_{1\le i< j\le 4 \atop {(ij)\neq (kl)}} 
              \frac{(-1)^{n_{ij}}\sqrt{2n_{ij}+1}}{B_{n_{ij}}}(\rho_{ij})^{n_{ij}}.
\label{diff.Ps}
\end{align}
Similarly ${\cal D}^1(x){\cal D}^1(y)G(\{\rho_{ij}+x\tau_{ij}+y\zeta_{ij}\};l_1l_2l_3l_4;L)$ is obtained from
\begin{align}
&{\cal D}^1(x){\cal D}^1(y)G(\{\rho_{ij}+x\tau_{ij}+y\zeta_{ij}\};l_1l_2l_3l_4;L)\nonumber \\
&\quad =\sum_{n_{12}n_{13}n_{14}n_{23}n_{24}n_{34}}{\cal D}^1(x){\cal D}^1(y) Ps(\{n_{ij}\};\{\rho_{ij}+x\tau_{ij}+y\zeta_{ij}\})\, Z(\{n_{ij}\};L) \nonumber \\
&\quad \ \ \times \delta(n_{12}+n_{13}+n_{14},l_1)\, \delta(n_{12}+n_{23}+n_{24},l_2)\, \delta( n_{13}+n_{23}+n_{34},l_3)\, \delta( n_{14}+n_{24}+n_{34},l_4).
\end{align}
Here, ${\cal D}^1(x){\cal D}^1(y) Ps(\{n_{ij}\};\{\rho_{ij}+x\tau_{ij}+y\zeta_{ij}\})$ is given by 
\begin{align}
&{\cal D}^1(x){\cal D}^1(y) Ps(\{n_{ij}\};\{\rho_{ij}+x\tau_{ij}+y\zeta_{ij}\})\nonumber \\
&\quad =
{\cal D}^1(y) \sum_{1\le k< l\le 4} \frac{(-1)^{n_{kl}}\sqrt{2n_{kl}+1}}{B_{n_{kl}}}
               n_{kl}\tau_{kl}(\rho_{kl}+y\zeta_{kl})^{n_{kl}-1}
              \prod_{1\le i< j\le 4 \atop {(ij)\neq (kl)}} 
              \frac{(-1)^{n_{ij}}\sqrt{2n_{ij}+1}}{B_{n_{ij}}}(\rho_{ij}+y\zeta_{ij})^{n_{ij}}\nonumber \\
&\quad =\sum_{1\le k< l\le 4} \frac{(-1)^{n_{kl}}\sqrt{2n_{kl}+1}}{B_{n_{kl}}}
               n_{kl}(n_{kl}-1)\tau_{kl}\zeta_{kl}(\rho_{kl})^{n_{kl}-2}
              \prod_{1\le i< j\le 4  \atop {(ij)\neq (kl)}} 
              \frac{(-1)^{n_{ij}}\sqrt{2n_{ij}+1}}{B_{n_{ij}}}(\rho_{ij})^{n_{ij}}\nonumber \\
&\quad \ \ + \sum_{1\le k< l\le 4,\   1\le k' < l' \le 4 \atop (kl) \ne (k'l')}
\frac{(-1)^{n_{kl}}\sqrt{2n_{kl}+1}}{B_{n_{kl}}}n_{kl}\tau_{kl}(\rho_{kl})^{n_{kl}-1}
\frac{(-1)^{n_{k'l'}}\sqrt{2n_{k'l'}+1}}{B_{n_{k'l'}}}n_{k'l'}\zeta_{k'l'}(\rho_{k'l'})^{n_{k'l'}-1}\nonumber \\
&\qquad \qquad \qquad \qquad \times \prod_{1\le i< j\le 4 \atop {(ij)\neq (kl),\, (k'l')}} 
              \frac{(-1)^{n_{ij}}\sqrt{2n_{ij}+1}}{B_{n_{ij}}}(\rho_{ij})^{n_{ij}}.
\label{DxDyonP}
\end{align}

\begin{flalign}
&{\rm (VI)}\quad
\prod_{i=1}^4{\cal D}^{l_i}(\alpha_i){\cal Y}_{\kappa \mu}(\textstyle{\sum_{i=1}^4\gamma_i\alpha_i{\bm e}_i})\nonumber &\\
& \qquad \quad    \Rightarrow (4\pi)^{\frac{3}{2}} \sqrt{(2\kappa+1)!}\, \delta(\textstyle{\sum_{i=1}^4l_i,\kappa}) \Big(\prod_{i=1}^4\frac{(\gamma_i)^{l_i}}
{\sqrt{(2l_i+1)!}}\Big)
[Y_{l_1}(\bm e_1)\!\times \!Y_{l_2}(\bm e_2)\!\times \!Y_{l_3}(\bm e_3)\!\times \!Y_{l_4}(\bm e_4)]_{\kappa \mu}.
\end{flalign}
\noindent
{\it Proof.\ \ }  It is easy to confirm (VI) by using successively the identity, Eq.~(\ref{Y.exp}). Note that, because $\sum_{i=1}^4l_i=\kappa$, 
\begin{align}
[Y_{l_1}(\bm e_1)\!\times \!Y_{l_2}(\bm e_2)\!\times \!Y_{l_3}(\bm e_3)\!\times \!Y_{l_4}(\bm e_4)]_{\kappa \mu}=
\Big[[Y_{l_1}(\bm e_1)\!\times \!Y_{l_2}(\bm e_2)]_{l_1+l_2} \times [Y_{l_3}(\bm e_3)\!\times \!Y_{l_4}(\bm e_4)]_{l_3+l_4}\Big]_{\kappa \mu}.
\end{align}

\begin{flalign}
&{\rm (VII)}\quad {\rm If}\  a+b+c+d=\kappa, \  {\rm then} \nonumber &\\ 
&\qquad \ \ \ \int \cdots \int d{\bm e}_1 \cdots d{\bm e}_4  [Y_{l_3}({\bm e}_3)\!\times \! Y_{l_4}({\bm e}_4)]^*_{L'M'} [Y_{l_1}({\bm e}_1)\!\times \! Y_{l_2}({\bm e}_2)]_{LM} \nonumber &\\
&\qquad \ \ \ \  \times 
[Y_{a}(\bm e_1)\!\times \!Y_{b}(\bm e_2)\!\times \!Y_{c}(\bm e_3)\!\times \! Y_{d}(\bm e_4)]_{\kappa \mu}\, \Big[[Y_{l_1-a}({\bm e_1})\!\times \!Y_{l_2-b}(\bm e_2)]_{\bar L} \! \times \!  [Y_{l_3-c}({\bm e_3})\! \times \! Y_{l_4-d}(\bm e_4)]_{\bar L}\Big]_{00}\nonumber &\\
& \qquad \ \ \ \Rightarrow (-1)^{l_1+l_2}\sqrt{\frac{2\kappa+1}{2L'+1}}\langle L\, M\ \kappa\, \mu|L'\, M'\rangle\,  \Lambda(abcd,l_1l_2l_3l_4, \bar{L};\kappa  L L'),\nonumber &\\
& \qquad \ {\rm where}\nonumber &\\
& \qquad \qquad \Lambda(abcd,l_1l_2l_3l_4,\bar{L}; \kappa L L')=
C(a \ l_1\!-\!a;l_1)\, C(b\ l_2\!-\!b;l_2)
 C(c \ l_3\!-\!c;l_3)\, C(d \ l_4\!-\!d;l_4) \nonumber &\\
& \qquad \qquad  \qquad  \qquad  \qquad  \qquad  \qquad \ \ \ \times 
\left[
\begin{array}{ccc}
\bar{L} & \bar{L} & 0 \\
a+b & c+d & \kappa \\
L & L' & \kappa\\
\end{array}
\right] 
\left[
\begin{array}{ccc}
a & b & a\!+\!b \\
l_1\!-\!a & l_2\!-\!b & \bar{L} \\
l_1& l_2 & L\\
\end{array}
\right]
\left[
\begin{array}{ccc}
c & d & c\!+\!d \\
l_3\!-\!c & l_4\!-\!d & \bar{L} \\
l_3& l_4 & L'\\
\end{array}
\right].&
\label{angle.integration}
\end{flalign}
\noindent
{\it Proof.\ \ } 
Since $a+b+c+d=\kappa$, $[Y_{a}(\bm e_1)\!\times \!Y_{b}(\bm e_2)\!\times \!Y_{c}(\bm e_3)\!\times \! Y_{d}(\bm e_4)]_{\kappa \mu} =   
\Big[[Y_{a}(\bm e_1)\!\times \!Y_{b}(\bm e_2)]_{a+b}\ \times \ [Y_{c}(\bm e_3)\!\times \! Y_{d}(\bm e_4)]_{c+d}\Big]_{\kappa \mu}$. 
Then it follows that 
\begin{align}
&[Y_{a}(\bm e_1)\!\times \!Y_{b}(\bm e_2)\!\times \!Y_{c}(\bm e_3)\!\times \! Y_{d}(\bm e_4)]_{\kappa \mu}\, \Big[[Y_{l_1-a}({\bm e_1})\!\times \!Y_{l_2-b}(\bm e_2)]_{\bar L} \! \times \!  [Y_{l_3-c}({\bm e_3})\! \times \! Y_{l_4-d}(\bm e_4)]_{\bar L}\Big]_{00}\nonumber \\
& \ =\sum_{L,L'} 
\left[
\begin{array}{ccc}
a+b & c+d & \kappa \\
\bar{L} & \bar{L} & 0 \\
L & L' & \kappa\\
\end{array}
\right]
\Big[ \big[[Y_{a}(\bm e_1)\!\times \!Y_{b}(\bm e_2)]_{a+b}\ \times \ [Y_{l_1-a}({\bm e_1})\!\times \!Y_{l_2-b}(\bm e_2)]_{\bar L} \big]_{L}\nonumber \\
& \qquad \qquad  \qquad \qquad \qquad \ \  \times \big[ [Y_{c}(\bm e_3)\!\times \! Y_{d}(\bm e_4)]_{c+d}\ \times \ [Y_{l_3-c}({\bm e_3})\! \times \! Y_{l_4-d}(\bm e_4)]_{\bar L}\big]_{L'}\Big]_{\kappa \mu}.
\end{align}
Because of the stretched coupling I obtain 
\begin{align}
&\big[[Y_{a}(\bm e_1)\!\times \!Y_{b}(\bm e_2)]_{a+b}\ \times \ [Y_{l_1-a}({\bm e_1})\!\times \!Y_{l_2-b}(\bm e_2)]_{\bar L} \big]_{L}\nonumber \\
& \quad \Rightarrow
\left[
\begin{array}{ccc}
a & b & a\!+\!b \\
l_1\!-\!a & l_2\!-\!b & \bar{L} \\
l_1& l_2 & L\\
\end{array}
\right] C(a \ l_1\!-\!a;l_1)\, C(b\ l_2\!-\!b;l_2)
[Y_{l_1}({\bm e}_1)\!\times \! Y_{l_2}({\bm e}_2)]_{L},\nonumber \\
&\big[ [Y_{c}(\bm e_3)\!\times \! Y_{d}(\bm e_4)]_{c+d}\ \times \ [Y_{l_3-c}({\bm e_3})\! \times \! Y_{l_4-d}(\bm e_4)]_{\bar L}\big]_{L'}\nonumber \\
& \quad \Rightarrow
\left[
\begin{array}{ccc}
c & d & c\!+\!d \\
l_3\!-\!c & l_4\!-\!d & \bar{L} \\
l_3& l_4 & L'\\
\end{array}
\right] C(c \ l_3\!-\!c;l_3)\, C(d \ l_4\!-\!d;l_4)
[Y_{l_3}({\bm e}_3)\!\times \! Y_{l_4}({\bm e}_4)]_{L'}.
\end{align}
The integral to be computed is given by 
\begin{align}
&\int \cdots \int d{\bm e}_1 \cdots d{\bm e}_4  [Y_{l_3}({\bm e}_3)\!\times \! Y_{l_4}({\bm e}_4)]^*_{L'M'} [Y_{l_1}({\bm e}_1)\!\times \! Y_{l_2}({\bm e}_2)]_{LM}
\Big[[Y_{l_1}({\bm e}_1)\!\times \! Y_{l_2}({\bm e}_2)]_{L}\ \times \ 
[Y_{l_3}({\bm e}_3)\!\times \! Y_{l_4}({\bm e}_4)]_{L'}\Big]_{\kappa \mu}\nonumber \\
&=(-1)^{l_1+l_2-L+M}\langle L\, -\!M\ L'\, M'|\kappa \mu \rangle=(-1)^{l_1+l_2+L+L'+\kappa}\sqrt{\frac{2\kappa+1}{2L'+1}}\langle L\, M\ \kappa\, \mu|L'\, M'\rangle. 
\end{align}
Combining the above equations and using the symmetry property of the $9j$ symbol
\begin{align}
\left[
\begin{array}{ccc}
a+b & c+d & \kappa \\
\bar{L} & \bar{L} & 0 \\
L & L' & \kappa\\
\end{array}
\right]=(-1)^{a+b+c+d+2\bar{L}+L+L'+2\kappa}
\left[
\begin{array}{ccc}
\bar{L} & \bar{L} & 0 \\
a+b & c+d & \kappa \\
L & L' & \kappa\\
\end{array}
\right],
\end{align}
I obtain $\Lambda(abcd,l_1l_2l_3l_4, \bar{L}; \kappa L L')$ as given above.

\acknowledgments
I am very much grateful to D. Baye and P. Descouvemont for valuable communications and 
to W. Horiuchi and M. Kimura for several discussions.

\end{document}